\documentclass[final]{aipproc}
\usepackage{graphicx}
\usepackage{amssymb}
\usepackage{amsmath}
\usepackage{color}
\usepackage[rflt]{floatflt}
\usepackage{mathptm}
\usepackage{epstopdf}
\layoutstyle{6x9}

\renewcommand{\theequation}{\arabic{equation}}

\newcommand{\pvalt}{\raise0.15ex\hbox{-}\mkern-11.5mu\int}
\newcommand{\be}{\begin{equation}}
\newcommand{\ee}{\end{equation}}
\newcommand{\bea}{\begin{eqnarray}}
\newcommand{\eea}{\end{eqnarray}}
\newcommand{\ben}{\begin{enumerate}}
\newcommand{\een}{\end{enumerate}}
\newcommand{\bit}{\begin{itemize}}
\newcommand{\eit}{\end{itemize}}

\newcommand{\half}{\frac{1}{2}}

\newcommand{\p}{\partial}

\def\fracs#1#2{\ensuremath{{\textstyle\frac{#1}{#2}}}}

\newcommand{\beq}{\begin{equation}}
\newcommand{\eeq}{\end{equation}}
\newcommand{\ba}{\begin{array}}
\newcommand{\ea}{\end{array}}

\newcommand{\<}{\langle}
\renewcommand{\>}{\rangle} 

\newcommand{\partslash}{\partial\hspace*{-5.5pt}\slash\hspace{2pt}}
\newcommand{\Dslash}{D\hspace*{-7pt}\slash\hspace{2pt}}

\definecolor{BrickRed}{cmyk}{0,0.89,0.94,0.28}
\definecolor{MidnightBlue}{cmyk}{0.98,0.13,0,0.43}
\definecolor{DarkGreen}{rgb}{0,0.7,0.1}

\newcommand{\rjcomm}[1]{{}}
\newcommand{\rjdel}[1]{{}}

\newcommand{\pdcomm}[1]{{}}
\newcommand{\pddel}[1]{}

\newcommand{\asdel}[1]{}
\newcommand{\ascom}[1]{}
\begin{document}


\title{Parity Doubling Among the Baryons}

\author{R.~L.~Jaffe, D.~Pirjol, and A.~Scardicchio}{
address={Center for Theoretical Physics,\\
Laboratory for Nuclear Science and Department of Physics\\
Massachusetts Institute of Technology, \\Cambridge, Massachusetts
02139}, } \rightline{MIT-CTP-3719}

\begin{abstract}
We study the evidence for and possible origins of parity doubling
among the baryons.    First we explore the experimental evidence,
finding a significant signal for parity doubling in the
non-strange baryons, but little evidence among strange baryons.
Next we discuss potential explanations for this phenomenon.
Possibilities include suppression of the violation of the flavor
singlet axial symmetry ($U(1)_{A}$) of QCD, which is broken by the
triangle anomaly and by quark masses.  A conventional Wigner-Weyl
realization of the $SU(2)_{L}\times SU(2)_{R}$ chiral symmetry
would also result in parity doubling. However this requires the
suppression of families of \emph{chirally invariant} operators by
some other dynamical mechanism. In this scenario the parity
doubled states should decouple from pions. We discuss other
explanations including connections to chiral invariant short
distance physics motivated by large $N_{c}$ arguments as suggested
by Shifman  and others, and intrinsic deformation of relatively
rigid highly excited hadrons, leading to parity doubling on the
leading Regge trajectory. Finally we review the spectroscopic
consequences of chiral symmetry using a formalism introduced by
Weinberg, and use it to describe two baryons of opposite parity.
\end{abstract}

%
%
\maketitle

%
\section{Introduction}

The possibility that excited hadrons occur in nearly degenerate
pairs of opposite parity ---  a phenomenon known as parity
doubling --- has been considered from time to time since the
1960's. Early arguments were based on Regge theory~\cite{regge}.
There have been speculations that dynamical symmetries, not
appearing in the underlying QCD Lagrangian, might be responsible
for this phenomenon\cite{iachello,Kirchbach:1997zb}.  Most
recently it has been suggested that parity doubling can be
explained by ``restoration'' of the underlying $SU(2)_{L}\times
SU(2)_{R}$ chiral symmetry of QCD at high energies
\cite{glozman1,glozman2,review,Inopin:2004iy,shifman} and/or in
specific sectors of the spectrum\cite{Jido:1999hd,Jido:2001nt}.

In this paper we carry out a detailed analysis, both
phenomenological and theoretical, of parity doubling among the
baryons made of $u$, $d$, and $s$ quarks. We do not consider
mesons here because there is less data and more controversy. Given
a concordance on meson resonances, our methods could easily be
applied.

We address three questions:  First, what is the experimental
evidence for parity doubling?   Second, could a symmetry of the
underlying QCD Lagrangian be responsible for parity doubling?  And
third, even if the origins of parity doubling lie elsewhere,
$SU(2)_{L}\times SU(2)_{R}$  is of fundamental importance for QCD,
and is widely discussed in connection with spectroscopy.  What, if
any, are its implications for the classification of hadrons? On
the first question, we find significant evidence for parity
doubling among non-strange baryons (nucleons and $\Delta$s), but
only weak evidence among hyperons ($\Sigma$s and $\Lambda$s) with
strangeness ($S$) minus one. There is not enough data to carry out
an analysis for $\Xi$s ($S=-2$) or $\Omega^{-}$s ($S=-3$). Better
information on the existence and classification of baryon
resonances would help significantly here.

Turning to the possible origins of parity doubling, the most
elegant explanations proposed make use of the fundamental
symmetries of the QCD Lagrangian: chiral $SU(2)_L \times SU(2)_R$
and the singlet axial symmetry $U(1)_A$. Any underlying symmetry
behind parity doubling must involve charges that transform as
pseudoscalars, since the multiplets involve states of opposite
parity.  The natural candidates are the charges of the isospin
axial currents, $Q^{a}_{5}$ ($a=1,2,3$), and the flavor singlet
axial charge, $Q_{5}$. $U(1)_{A}$ connects states of opposite
parity, but the same flavor and spin.  $SU(2)_{L}\times SU(2)_{R}$
charges change both parity and isospin, so multiplets in general
contain states of opposite parity and different isospin.

Chiral symmetry restoration has been proposed as a mechanism for
parity doubling for high excitations in the baryons spectrum
\cite{glozman1,glozman2,review,Inopin:2004iy,shifman}. This
phenomenon was given the name ``effective chiral restoration'',
which we also use in the following for ease of comparison with the
literature. It is equivalent to realizing chiral symmetry linearly
on the excited states. As we pointed out in a recent
paper~\cite{us}, $SU(2)_{L}\times SU(2)_{R}$ chiral symmetry, when
applied to the effective Lagrangian describing the interactions of
excited baryons and matter, does not, by itself, lead to parity
doubling. One can write down what appear to be Wigner-Weyl
representations, where hadrons of equal mass and opposite parity
transform into one another under the action of the $SU(2)$ axial
charges.  However $SU(2)_{L}\times SU(2)_{R}$ \emph{invariant}
operators that couple the would-be parity doublets to pions can be
added to the effective Lagrangian describing these particles.
These operators have the effect of removing the degeneracy of the
previously degenerate states and spoiling any relationship between
their axial charges.  Non-linear field redefinitions make the
situation clear (and do not change the physics): The only
irreducible representations are multiplets of definite isospin and
parity that transform according to the non-linear
realization~\cite{Cheff,Weinberg:1968de,book,CCWZ}. To obtain
degenerate pairs of opposite parity (and hence to achieve parity
doubling) it is necessary to suppress the families of
$SU(2)_{L}\times SU(2)_{R}$ \emph{invariant} pion couplings that
split them.

The existence of these operators and the need to suppress them
dynamically  has not been systematically examined in work on
effective chiral symmetry restoration
\cite{glozman1,glozman2,review}.

A somewhat different approach is taken in
Refs.~\cite{Jido:1999hd,Jido:2001nt}, where hadrons are assigned
to specific representations of $SU(2)_{L}\times SU(2)_{R}$ in the
context of a linear sigma model.  Chiral symmetry is subsequently
broken when the sigma field takes a vacuum expectation value.  By
assigning hadrons to minimal (though not irreducible)
$SU(2)_{L}\times SU(2)_{R}$ representations, and allowing only low
dimension operators in the linear sigma model Lagrangian, a
phenomenological framework for describing baryon resonances is
constructed.  In some examples,
Refs.~\cite{Jido:1999hd,Jido:2001nt} take parity doubling as an
input when they choose representations and interactions.   In fact
the example we discuss at length in Sec.~III was first introduced
in Ref.~\cite{DeTar:1988kn} as a ``mirror'' model of opposite
parity baryons.  Because Refs.~\cite{Jido:1999hd,Jido:2001nt} do
not attempt either to marshall evidence for parity doubling in the
data or explain it in terms of some more fundamental dynamics, we
do not discuss them further here.

If ``effective chiral symmetry restoration'' is responsible for
parity doubling, then states, to the extent they are parity
doubled, must decouple from pions.  This prediction can be tested
experimentally.  Furthermore, only the simplest $SU(2)_{L}\times
SU(2)_{R}$ multiplets involve only parity doubled pairs of states.
Generic representations include states of different isospin as
well as parity, which should be characteristic of this mechanism
for parity doubling.

In contrast, the $U(1)_{A}$ symmetry of QCD is broken explicitly
by a term in the quantum action induced by instantons
\cite{thooft} and, of course, by quark masses. Thus, its
generator, $Q_5$, the associated singlet axial charge, does not
commute with the QCD Hamiltonian\cite{thooft}.  It is also not a
symmetry of the QCD vacuum, since, as Coleman showed long
ago\cite{coleman}, a charge that does not commute with the
Hamiltonian cannot annihilate the vacuum.  Thus $Q_{5}|0\>\ne 0$.
Note, however, that this \emph{does not} mean that $U(1)_{A}$ is
``spontaneously broken''. In a theory like QCD with a
non-perturbative $\<\bar qq\>$ condensate in addition to
instantons and quark masses, the question of whether a chiral
symmetry is explicitly or spontaneously broken is \emph{a matter
of degree}.  At one extreme, for zero up and down quark mass,
$SU(2)_{L}\times SU(2)_{R}$ is clearly spontaneously broken. As an
example of the other extreme, consider QED:  Since $m_{e}\ne0$,
the axial $U(1)_{A}$ associated with the current $\bar
e\gamma_{\mu}\gamma_{5}e$ is not a symmetry of QED. Its charge
does not commute with the Hamiltonian and does not annihilate the
vacuum.  In fact one can compute the $\bar e e$ condensate in
perturbation theory. Specifically, in the $\overline{MS}$ scheme
this is $\langle\overline{e} e\rangle_{\mu} = m^3/(4\pi^2)
f(\alpha)$, where $f(\alpha)=  1 + \ell + \alpha/(2\pi)(5 + 5 \ell
+ 3\ell^2) + O(\alpha^2)$,
 with $\ell = \log(\mu^2/m^2)$
and $\mu$ is the renormalization point \cite{qedvev}.  Despite the
fact that $\<\overline e e\>\ne 0$, we say this symmetry is
\emph{explicitly broken}.  In QCD with three colors, the
importance of \emph{explicit} $U(1)_{A}$ symmetry violation can be
gauged by the magnitude of the $\eta'$ mass, which comes entirely
from explicit symmetry violation and is much greater than the
scale of spontaneous symmetry breaking, $f_{\pi}$. Therefore, as
discussed by 't Hooft\cite{thooft}, $U(1)_{A}$ behaves like an
explicitly broken symmetry. This is one place where it is
\emph{not} useful to take the large $N_{c}$ limit in QCD.  There
the situation becomes reverses because the source of explicit
$U(1)_{A}$ symmetry violation disappears and the $\eta'$ becomes a
ninth pseudoscalar Goldstone boson (when $m_{u,d,s}=0$). In that
limit the $U(1)_{A}$ symmetry should be regarded as spontaneously
broken. However, we live in a world with $N_{c}=3$ and explicit
breaking of $U(1)_{A}$ dominates.

Like any \emph{explicitly broken} symmetry, if the matrix elements
\emph{that violate $U(1)_{A}$ symmetry} are small  in a sector of
the spectrum, some of its consequences will follow.  In the case
at hand, parity doubling will be seen in a sector of the spectrum
if the matrix elements of the divergence of the flavor singlet
axial current are dynamically suppressed in that sector
\cite{duck}.  Other consequences of the symmetry are not obtained:
for example the parity doubled states do not necessarily form
representations of $U(1)_{A}$.  For the sake of clarity, we refer
to this phenomenon as ``dynamical suppression of $U(1)_{A}$
symmetry violation'' rather than ``$U(1)_{A}$ symmetry
restoration''.\footnote{In an earlier version of this paper we
argued that suppression of $U(1)_{A}$ violation would be less
likely among strange baryons.  We no longer believe this to be
required.  We thank A.~Manohar for discussions on this point.}

To summarize, either chiral symmetry or $U(1)_A$ could be
responsible for parity doubling. In both cases extra dynamical
mechanisms must be at work.  For $SU(2)_{L}\times SU(2)_{R}$,
\emph{chirally invariant} operators must be suppressed. If so,
hadrons form chiral multiplets, not necessarily limited to parity
doublets, and they decouple from pions. $U(1)_{A}$ is more
conventional:  parity doubling can result if operators that
\emph{violate the symmetry} are suppressed.

We also briefly explore some other proposed dynamical origins for
parity doubling.   We consider the possibility that parity
doubling is not directly associated with a symmetry  of the
underlying Lagrangian, but instead  is a consequence of an
intrinsic deformation of excited hadrons. If baryons can be
accurately described as intrinsically deformed systems, if the
intrinsic state spontaneously violates parity, and if the
intrinsic state is rigid, {\it i.e.\/} the matrix element between
the state and its parity image is small, then when collectively
quantized, the spectrum would exhibit parity doubling. This is, in
fact, an old idea\cite{regge,iachello,Johnson:1975sg}, developed
in early geometrical models of baryons. The arguments in support
of this mechanism seem to apply best to states of high angular
momentum, and there is evidence for parity doubling at high $J$
among non-strange baryons.  However we also find strong evidence
for parity doubling among baryons with $J=1/2$ and $3/2$, which is
hard to motivate from this point of view. We also discuss the
arguments given by Shifman~\cite{shifman} that the chiral
invariance of perturbative QCD, valid at short distances, together
with certain dynamical assumptions motivated by the $N_{c}\to
\infty$ ($N_{c}$ is the number of colors in QCD) limit, lead to
parity doubling among mesons, and perhaps by extension, baryons.

Finally, although $SU(2)_{L}\times SU(2)_{R}$ is in general not
apparent in the spectrum, it is not devoid of implications for the
spectrum of baryons.  Weinberg has shown that hadrons can be
classified into multiplets under an  $SU(2)_{L}\times SU(2)_{R}$
symmetry where the axial generators are associated with
pion-hadron vertices, provided hadron-hadron scattering amplitudes
obey certain constraints at asymptotic energies
\cite{Weinberg:1969hw,Weinberg:1990xn}. Weinberg's ``mended''
chiral symmetry does not naturally lead to parity doubling,  but
it can accommodate it for special choices of the hadrons'
assignments into $SU(2)_{L}\times SU(2)_{R}$ representations. We
explore the possible spectroscopic consequences of chiral symmetry
in the context of Weinberg's work on ``mended symmetries'', and
comment on the relation of this approach to the contracted
spin-flavor symmetry emerging in the large $N_c$ limit \cite{DM,
DJM}.

The remainder of the paper is organized as follows. In the next
section we examine the experimental situation. Despite all the
theoretical work, we know of no systematic attempt to quantify the
evidence for parity doubling. Typically papers on the subject
include  graphs or tables where the reader is invited to see some
evidence for a correlation between states of opposite parity (for
a recent example, see Ref.~\cite{Inopin:2004iy}).  It is
difficult, however, to factor in the (often large) widths of
states and the fact that some states are much better established
than others. Also it is not clear which state to pair with which.
Some previous attempts to quantify the evidence for (or against)
parity doubling can be found in
Refs.~\cite{Glozman:2004gk,Klempt:2002tt}.

We introduce a measure of the correlation between states of
opposite parity, $\Omega(I,J,S)$, that has some statistical basis.
Specifically, we test the hypothesis that two functions, positive
and negative parity spectral densities for each isospin ($I$),
strangeness ($S$), and spin ($J$), are identical within a
tolerance $\sigma$, which measures the extent of symmetry
breaking.  We construct the spectral functions for each $I$ and
$J$ from the baryons accepted by the Particle Data
Group~\cite{PDG}, including widths, and weighting resonances to
reflect their reliability.  We have no {\it a priori\/} standard
with which to compare Nature's value of $\Omega(I,J,S)$.  Instead
we construct a set of ``alternative realities'' by scrambling the
parities of the known states.  As an ensemble, this set is free
from correlations, so the comparison between $\Omega(I,J,S)$
computed over this ensemble with the value obtained from Nature's
assignment of parities gives a measure of the significance of the
correlation.  The method and our results are discussed in detail
in Section II.

In Section III we look at the fundamental symmetries of QCD to see
which, if any, could be responsible for the phenomenon of parity
doubling. We review the realizations of chiral symmetry and
briefly summarize  the arguments of our recent paper \cite{us} on
the way that massless Goldstone bosons transforming non-linearly
undermine the predictions of linearly realized chiral symmetry for
baryon masses and couplings. We go on to state the conditions on
the chiral effective Lagrangian that must be satisfied in order to
obtain parity doubling, irrespective of the underlying dynamical
mechanism responsible for it. We describe the phenomenon of
``dynamical suppression of $U(1)_A$ breaking'', and show that it
can lead to parity doubling in a sector of the Hilbert space of
QCD. We also summarize and review arguments for parity doubling
from chiral symmetry restoration at short distances, and due to
intrinsic hadron deformation. For completeness we summarize the
spectroscopic predictions of Weinberg's approach in Section IV,
where we introduce a new example that reproduces naturally some of
the predictions of $SU(2)_{L}\times SU(2)_{R}$ restoration,
without fine tuning of the coefficients in the chiral Lagrangian.
Finally in Section  V we summarize our results and discuss areas
for further work.

\section{What is the Experimental Evidence?}

In this section we evaluate the experimental evidence for parity
doubling among the baryons.  We have chosen to study the baryons
as listed by the Particle Data Group\cite{PDG}.  There is still
much uncertainty concerning the spectrum of light ($u$, $d$, $s$)
baryons and mesons, and other authors have chosen to study
different data\cite{glozman1,glozman2}.  We chose the PDG listings
because they provide a critical summary of many experiments over
many decades, and also because they treat baryon states uniformly,
assigning masses, widths, and ``reliabilities'' using their famous
``star'' system.  Data on the meson spectrum accepted by the PDG
is less complete, more controversial and lacks the reliability
assignments. This point requires some discussion.  The PDG has
assigned reliabilities to 44 non-strange baryon
resonances\cite{baryontable}.  For each angular momentum there are
\emph{two} isospin channels, $I=1/2$ (nucleon) and $I=3/2$
($\Delta$).  The PDG non-strange meson listings are harder to
interpret.  The ``Meson Summary Table'' lists (coincidently) 44
well-established non-strange mesons. However those states are
distributed over \emph{four} channels, $I=0$ and $I=1$ with $C=\pm
1$.  Many more states appear in the full meson listings:  70
states can be found in the non-strange meson listings and another
96(!) can be found under the heading ``Other Light Unflavored
Mesons''.  The latter list includes some that have been seen in
only one experiment and some that other groups have claimed to
exclude.  Recently, Bugg\cite{Bugg:2004xu} has made an independent
analysis of the meson spectrum including an attempt at assigning
reliabilities, with results quite different from the PDG analysis.
It would be very interesting to repeat our statistical analysis on
a well-defined set of meson resonances.

\begin{figure}[ht]
\label{pointsND}
\includegraphics[width=7cm]{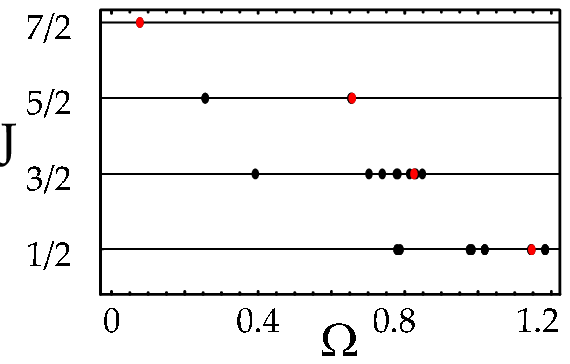}
\includegraphics[width=7cm]{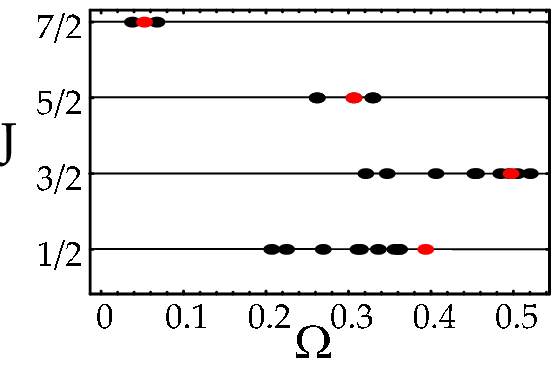}
\caption{\small The value of $\Omega$ at $\sigma=125$MeV for
nucleons (left) and $\Delta$'s (right) of different spins $J$. The
value of $\Omega$ with Nature's parity assignments  is marked with
a red circle.}
\end{figure}

\begin{figure}[ht]
\label{pointsSL}
\includegraphics[width=6.5cm]{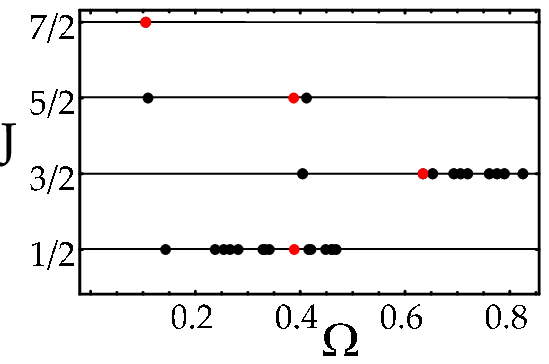}
\includegraphics[width=7cm]{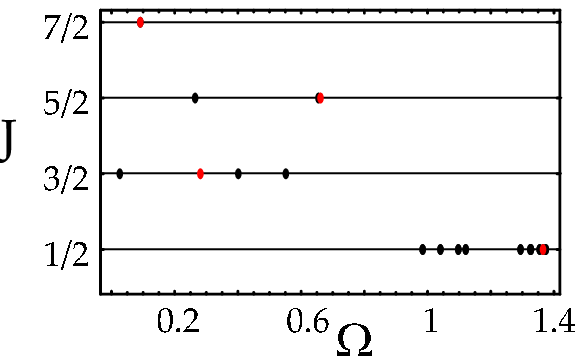}
\caption{\small The value of $\Omega$ at $\sigma=125$MeV for
$\Sigma$'s (left) and $\Lambda$'s (right) of different spins $J$.
The value of $\Omega$ with Nature's parity assignments is marked
with a red circle.}
\end{figure}

\begin{figure}[ht]
\label{deltatotal}
\includegraphics[width=7cm]{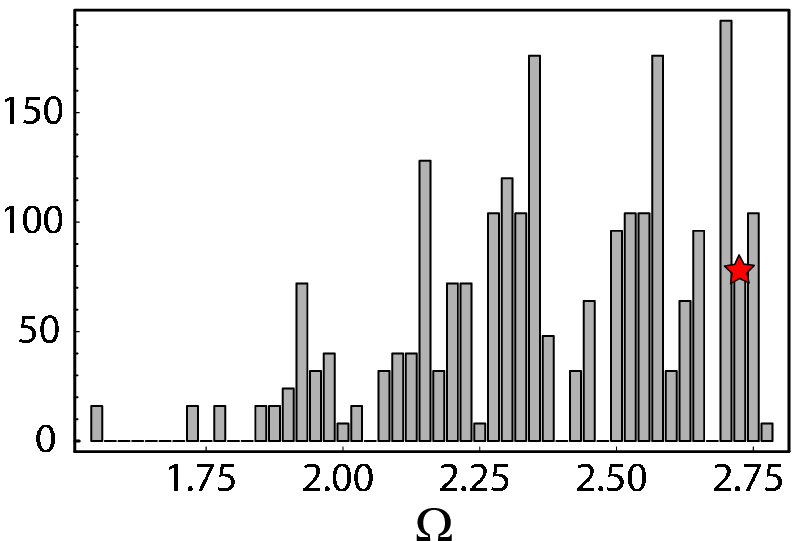}\includegraphics[width=7cm]{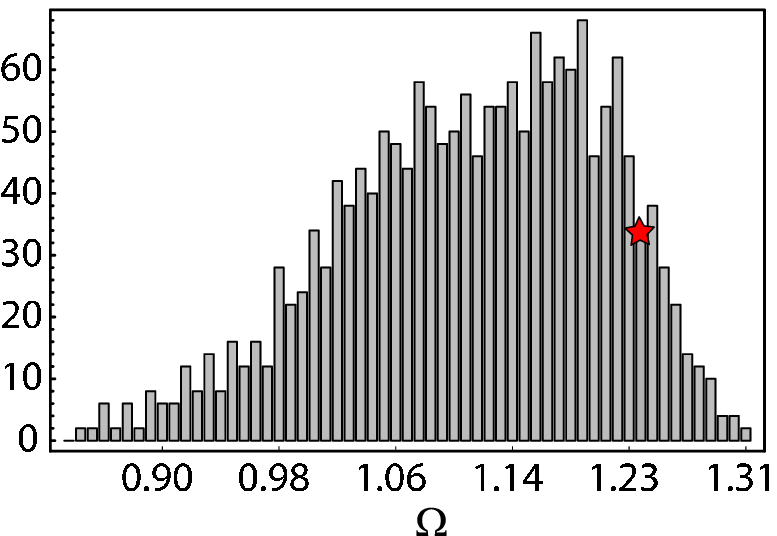}
\caption{\small Histograms of frequency of the value of $\Omega$
in all the possible realizations of parity assignments in the the
nucleons (left) and $\Delta$'s (right). All spins less than or
equal $\frac72$ are included. The value of $\Omega$ with the
parities as observed in Nature is indicated with a  star. The area
on the left of the bar represents the `confidence level' (C.L.)
attached to the given $\Omega$. For nucleons we have $95\%$ C.L.\
and for the $\Delta$'s $86\%$ C.L.\ }
\end{figure}

\begin{figure}[ht]
\label{sigmatotal}
\includegraphics[width=7.2cm]{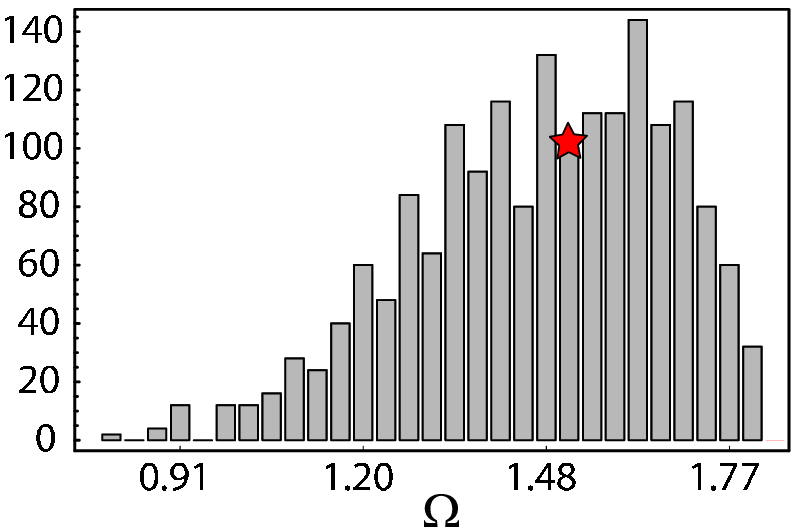}\includegraphics[width=6.9cm]{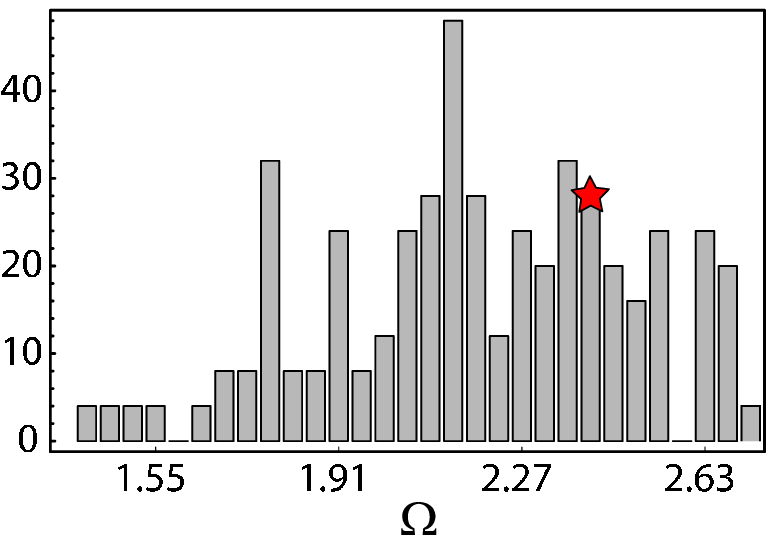}
\caption{\small Histograms of frequency of the value of $\Omega$
in all the possible realizations of parity assignements in the
$\Sigma$ (left) and $\Lambda$ (right). All spins $J=1/2,...,7/2$
are included. The value of $\Omega$ with the parities as observed
in Nature is indicated with   a star. The area on the left of the
bar represents the `confidence level' (C.L.) attached to the given
$\Omega$. For $\Sigma$ we have $60\%$ C.L.\ and for the $\Lambda$
$65\%$ C.L.\ .}
\end{figure}

We find that there is reasonably strong evidence for parity
doubling among the non-strange baryons ($I=1/2$ nucleons and
$I=3/2\ \Delta$s), and weaker, inconclusive evidence among the
$S=-1$ baryons ($I=0\ \Lambda$s and $I=1\ \Sigma$s). There is not
enough data on cascades ($S=-2$) or $\Omega$s ($S=-3$) to perform
our analysis. The results are summarized in Figs.~\ref{pointsND},
\ref{pointsSL}, \ref{deltatotal} and \ref{sigmatotal}, which
should be consulted after an explanation of our method.  It will
be clear that the study of parity doubling is hampered by our
incomplete knowledge of baryon resonances, especially those with
strangeness. Certainly a new experimental attack on hadron
spectroscopy could greatly improve matters.

Despite many papers on the subject over many years, we know of no
attempt to make a (semi-) quantitative evaluation of the evidence
for parity doubling.  Typically authors simply show the spectrum
of states with the same flavor and spin quantum numbers and
opposite parity, and rely on the eye of the reader to recognize a
propensity for states of opposite parity to cluster nearby in
mass.  Such an  ``eyeball'' approach has several shortcomings,
\begin{itemize}
\item Baryon resonances have large widths that are hard to display
graphically.

\item Different resonances have different statistical
significance.  Overlap of two well established resonances should
be given more weight than overlap of two doubtful states.

\item  There is no reason to expect parity doubling to be an exact
symmetry, so spectra must be compared with an allowance for
symmetry breaking.

\item Because of symmetry breaking and resonance widths, in
channels with many states it is often not clear which states are
to be compared.
\end{itemize}
Recently Glozman and Klempt\cite{Glozman:2004gk,Klempt:2002tt}
have made more quantitative analyses, which  however   still
suffer from some of the shortcomings listed above. In our analysis
we will address all of these difficulties.

We know of no way to obtain a rigorous measure of the significance
of parity doubling in the physical spectrum, which perhaps
explains the absence of work in this direction.  However we do
believe it is possible to improve on the eyeball. To do this one
needs two ingredients: 1)  A measure of how much correlation there
is between negative and positive parity states; and 2) A control
set of ``alternative worlds'' on which the correlation can be
measured and compared with our own.

 A realistic approach to generating control spectra has to respect
several features of the baryon resonance spectrum:  a) states are
sparse at low mass, where there is no evidence for parity
doubling; b) the data in each channel show roughly equal numbers
of positive and negative parity states; c) reliabilities decrease
and widths increase  in general with resonance mass.  Reference
spectra that ignore these features would not give fair comparisons
with the real world. We do not know an algorithm that would
generate \emph{a priori} satisfactory comparison spectra free of
these problems.

Instead we have chosen to use the physical spectrum itself to
generate comparisons.  Specifically, in each channel of definite
$J$, $I$, and $S$, we keep the masses, widths, and reliabilities
fixed, and reshuffle the the parities, keeping the number of
positive and negative parity states fixed.  This method generates
comparison spectra which respect the characteristics listed in the
previous paragraph, but which, as an ensemble, is free from any
additional parity correlation beyond the desirable feature that
the number of states of opposite parities is roughly equal in each
channel.  Comparing the real world with these alternative realites
tests whether the correlation between states of opposite parity in
the real world beats a random ensemble with otherwise similar
spectroscopic features.

We consider ``channels'' of specific total angular momentum, $J$,
isospin, $I$, and strangeness, $S$.  In each channel we define
positive and negative parity spectral functions,
$\rho_{JIS}^{\pm}(m)$, which we take to be a sum over normalized
Breit-Wigner resonances with masses and widths taken from the
PDG\footnote{We do not include the quoted uncertainties in masses
and widths since they are generally smaller than the widths of the
states, and including them would complicate our analysis without
adding much value.}.  Specifically, we include only the states
listed in the ``Note on Nucleon and $\Delta$
Resonances''\cite{baryontable}. We weigh the resonances by a
confidence factor, $W_{j}$, proportional to the number of stars
designated by the PDG, because we believe more weight should be
given to correlations between well established states, 
\begin{eqnarray}
\label{spectral}
\rho_{IJS}^{+}(m,\{C\})&=&\sum_{j}\frac{W_{j}}{{2}\pi}\frac{\Gamma_{j}}{(m-m_{j})^{2}+\Gamma_{j}^{2}/4}C_{j}\nonumber\\
\rho_{IJS}^{-}(m,\{C\})&=&\sum_{j}\frac{W_{j}}{{2}\pi}\frac{\Gamma_{j}}{(m-m_{j})^{2}+\Gamma_{j}^{2}/4}(1-C_{j})
\end{eqnarray}
The sum ranges over \emph{all} states with a definite $J$, $I$,
and $S$, and $W_{j} = 1.0, 0.75, 0.50, 0.25$ for $4^{*}$, $3^{*}$,
$2^{*}$, and $1^{*}$ resonances respectively.  We have also
performed the analysis with an exponential weighting,  $W_{j} =
1.0, 0.50, 0.25, 0.125$ respectively.  The effect on our results
is negligible except for the  $\Sigma$s ($I=1$, $S=-1$), where the
shape of the distribution shifts significantly enough to warrant
comment (see discussion below).  The factor $C_{j}$ assigns a
parity to the $j^{\rm th}$ state: $C_{j} = +1$ for positive
parity, and $C_{j}=0$ for negative parity.  Of course, Nature
prescribes a parity for each state as a set of $C_{j}$'s. Notice
that the $\rho_{IJS}^{\pm}$ are normalized,
\begin{equation}
\int_{-\infty}^\infty dm \rho^{\pm}_{IJS}(m,\{C\})\equiv
N_{IJS}^{\pm}(\{C\})=\sum_{j}W_{j}\left\{\begin{matrix} C_{j} &
\hbox{for}\quad \rho^{+}\\ (1-C_{j}) & \hbox{for} \quad
\rho^{-}\end{matrix} \right.
\label{spectralnorm}
\end{equation}
so the norm measures the number of states weighted by our
confidence that they are real.

We do not expect parity doubling to be an exact symmetry of QCD.
Instead we expect it to be approximately valid, parameterized by
some symmetry breaking scale $\sigma$, which is expected to be of
order (say) 50--200 MeV.  Our aim, then, is to test the hypothesis
that the masses of positive and negative parity states are
compatible within fluctuations with a variance $\sigma$.  A simple
indicator we can use to test the above hypothesis is
\begin{equation}
\label{statistic}
\Omega_{IJS}(\{C\}) =\int dm_{1}dm_{2}\rho_{IJS}^{+}(m_{1},\{C\})\
{\rm
erfc}\left(\frac{|m_{1}-m_{2}|}{\sigma}\right)\rho_{IJS}^{-}(m_{2},\{C\})
\end{equation}
where ${\rm erfc}(z) =
\frac{2}{\sqrt{\pi}}\int_{z}^{\infty}dte^{-t^{2}}$ is the
complementary error function.  ${\rm erfc}(0)=1$ and ${\rm
erfc}(z)\to 0$ when $z\to\infty$. Eq.~(\ref{statistic}) then can
be interpreted as follows. For a given $m_1$ and $m_2$, ${\rm
erfc}(|m_{1}-m_{2}|/\sigma)$ is the probability that a gaussian
random number $\mu$ with average $m_1$ and variance $\sigma$ gives
a $|\mu-m_1|\geq |m_2-m_1|$, {\it i.e.\/} the probability that
$m_2$ is closer to $m_1$ than a randomly generated number (with
variance $\sigma$). We can interpret $\sigma$ as the order of
magnitude of the breaking of the symmetry (the argument is
symmetric in $m_2\leftrightarrow m_1$).  The possible values of
$m_{1}$ and $m_{2}$ are then distributed according to
$\rho^{\pm}_{IJS}(m_{1,2},\{C\})$, normalized not to unity, but to
$N^{\pm}_{IJS}(\{C\})$. $\Omega_{IJS}(\{C\})$ measures the
correlation of the spectral functions with a ``forgiveness
factor'' $\sigma$.  $\Omega$ is going to be larger in channels
with more states, greater reliability, and, of course, better
correlation between states of opposite parity.  Since ${\rm
erfc}(x)\leq 1$ the maximum possible value of $\Omega_{IJS}$
occurs if the typical separations $|m_i-m_j|$ and widths
$\Gamma_i$ are much smaller than the chosen $\sigma$. In that case
we can set ${\rm erfc}(|m_{1}-m_{2}|/\sigma)\approx 1$ for all
$m_{1}$ and $m_{2}$ and the two integrals decouple giving and
$\Omega_{IJS}^*\equiv{\rm
Max}[\Omega_{IJS}(\{C\})]=N^{+}_{IJS}(\{C\})N^{-}_{IJS}(\{C\})$.
Some other features of $\Omega_{IJS}$ are worth noting: There is
no penalty for the absence of parity doubling low in the spectrum.
A state like the nucleon  with no parity partner simply does not
contribute.  Any overlapping states of opposite parity (and the
same $IJS$) contribute to $\Omega_{IJS}$.  No choice of multiplet
assignments is needed.  The contribution of broad states is
insensitive to shifts correlations on scales small compared to
their widths.

We do not know how to judge the {\it a priori\/} significance of
the experimental value of $\Omega_{IJS}$ because the underlying
distributions $\rho_{IJS}^{\pm}$ are not normalized to unity, but
instead weighted by the number and reliability in that channel ---
features that we regard to be important in the analysis.  So we
are led to check the value of $\Omega$ realized in Nature against
an ensemble of control spectra generated by scrambling the
parities among the otherwise fixed states for each $J$, $I$, and
$S$.   The crucial feature of these control spectra is that the
parities are reassigned randomly.  So the comparison between
Nature and the control set frames the question whether the
parities of the states with a given $I,J,S$ are more correlated
than a random assignment of parities.

\begin{figure}[ht]
\label{variationsigma}
\includegraphics[width=10cm]{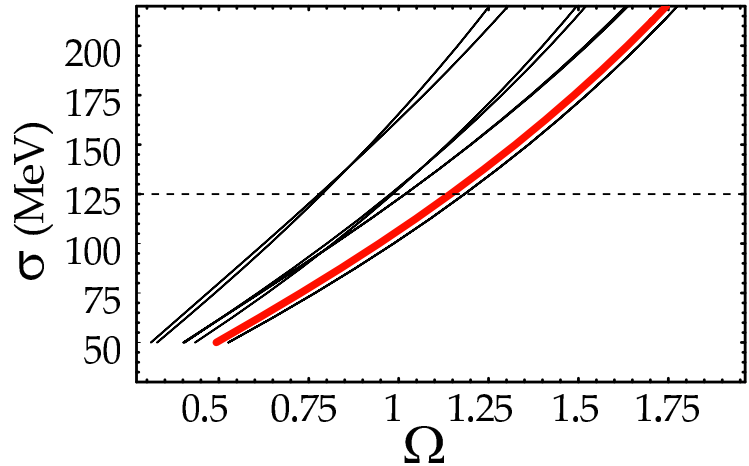}
\caption{\small The value of $\Omega$ for Nucleons ($I=1/2$,
$S=0$) with $J=1/2$ for varying $\sigma$. The value of $\Omega$
with the Nature's parity assignments  is the red line.}
\end{figure}

Fig.~\ref{variationsigma} shows the results in a typical channel,
nucleons with $J=1/2$.  $\Omega_{1/2,1/2,0}$ is plotted for
Nature's parity assignments (in red) and the control sample (in
gray), for a range of $\sigma$.  The ordering of instances turns
out not to depend significantly on $\sigma$, so we lose little by
choosing a fixed value of $\sigma$ for the rest of the analysis.
We have chosen $\sigma$ = 125 MeV, though we cannot attribute any
physical significance to this choice.  Fig.~\ref{pointsND}
presents the data for nucleons and $\Delta$s (for $J\le 7/2$) and
Fig.~\ref{pointsSL} shows the data for $\Sigma$s and $\Lambda$s.

Data are most abundant for nucleons and $\Delta$s, and for these
two channels Nature   shows typically more correlation than the
alternative realities. To give a quantitative answer, in terms of
confidence level, we construct a histogram of the frequency of a
given value of $\Omega$. Because the $\Omega_{IJS}$ is normalized
in a way that reflects the significance of each channel, it makes
sense to combine data for sets of channels, thereby obtaining a
larger control ensemble. This has been done in
Fig.~\ref{deltatotal} for nucleons and for $\Delta$s.  From these
figures it is clear that although a small number of parity
reassignments would do better, Nature's correlation between states
of opposite parity is quite striking. In numbers the `confidence
level' (the area on the left of the starred bin in Figure
\ref{deltatotal}) for the value of $\Omega$ in the Nature is
$95\%$ for the nucleons and $86\%$ for the $\Delta$s.

\begin{figure}[ht]
\label{newsigma}
\includegraphics[width=7cm]{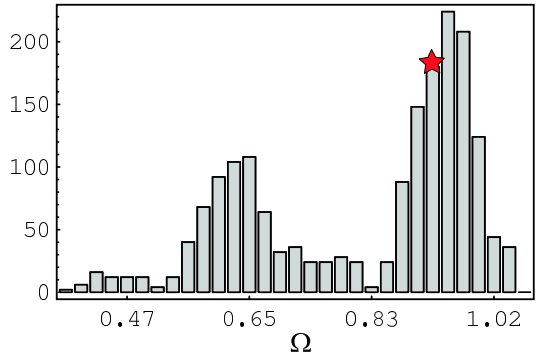}
\caption{\small Frequency histogram of the value of $\Omega$ for
the $\Sigma$'s using the ``exponential'' weighting of the
significance of states. Compare with Fig.~\ref{sigmatotal}, left
box.}
\end{figure}

The evidence for parity doubling among $\Lambda$s and $\Sigma$s,
channel by channel, is   shown in Fig.~\ref{pointsSL}, and the
statistic is summed over all $S=-1$ channels in
Fig.~\ref{sigmatotal}. For the strange baryons we find much
smaller values of the confidence level for $\Omega$: $58\%$ for
the $\Sigma$s and $72\%$ for the $\Lambda$s. We conclude that it
is not possible to make a definitive statement in this sector.
There may be some correlation, but more data are needed.
Fig.~\ref{newsigma} shows histogram using the ``exponential''
weighting factor for the $\Sigma$s.  Here, the change in weighting
makes a difference:  the signal for parity doubling is stronger
when well established resonances are more strongly weighted.

\section{Possible Origins of Parity Doubling}

In this section we examine some possible explanations for the
phenomenon of parity doubling among light quark baryons. In the
classical limit, the Lagrangian of QCD with $n_f$ light quark
flavors has the symmetry
\begin{eqnarray}
SU(n_f)_L \times SU(n_f)_R \times U(1)_A \times U(1)_V
\end{eqnarray}
This consists of the product of chiral symmetry, $SU(n_f)_L \times
SU(n_f)_R$, with the singlet axial symmetry, $U(1)_A$, and the
baryon number invariance $U(1)_V$. Our understanding of QCD is
profoundly shaped by these symmetries, so in the following we
review their possible role in connection with the phenomenon of
parity doubling. Since we are studying light quark spectroscopy
but wish to keep track of strange quark mass effects, we take
$n_{f}=2$.

We start by reviewing the implications of chiral symmetry for the
masses and couplings of hadrons. This part also introduces the
notation to be used in the remainder of the paper. We also briefly
review the arguments we presented in Ref.~\cite{us} that parity
doubling does not follow from the underlying $SU(2)_{L}\times
SU(2)_{R}$ symmetry of QCD.  That symmetry can only be restored in
the spectrum if classes of $SU(2)_{L}\times SU(2)_{R}$
\emph{invariant} operators are suppressed for some dynamical
reason.

Next we consider the possibility that parity doubling is a
consequence of dynamical suppression of flavor singlet axial
symmetry violation. As is usually the case with explicitly broken
symmetries, the symmetry can be approximately realized in the
spectrum if the matrix elements of operators that violate the
symmetry are suppressed.

We then discuss two other possibilities:  first Shifman's attempt
to relate parity doubling to chiral symmetry restoration at short
distances in QCD, and second the possibility that parity doubling
is associated with an intrinsic deformed baryon state, a mechanism
familiar from nuclear and molecular physics.

 \subsection{Effective restoration of $\mathbf{SU(2)_{L}\times SU(2)_{R}}$}

The Lagrangian of QCD with $n_f$ massless quarks has an exact
chiral SU$(n_f)_L \times $ SU$(n_f)_R$ symmetry, with generators
given by the vector $T^{a}$ and axial $X^{a}$ charges. The
generators satisfy the commutation relations
\begin{eqnarray}
[T^a, T^b] = if_{abc} T^c\,, \quad [T^a, X^b] = if_{abc} X^c\,,
\quad [X^a, X^b] = if_{abc} T^c\,,
\end{eqnarray}
where $a,b,c=1,...,n_{f}^{2}-1$.  For the case of $n_{f}=2$ of
most interest here, $f_{abc}=\varepsilon_{abc}$. The vector and
axial isospin currents are exactly conserved,
$\p^{\mu}j_{\mu}^{a}=\p^{\mu}j_{5\, \mu}^{a}=0$.  The generators
have different transformation properties under parity: the vector
generators, $T^a$, commute with parity, but the axial charges
anticommute $\{ X^a , P \} = 0$. This suggests that parity is
embedded nontrivially in representations of the chiral group, in a
way which could reproduce the observed parity doubling in the
hadron spectrum.

It is instructive at the outset to review the consequences of
current conservation for the spectrum.  Conservation of the vector
current requires hadrons to lie in degenerate multiplets of
definite isospin.  The consequences of axial current conservation
are embodied in generalized Goldberger-Treiman relations. Consider
two hadrons of identical flavor and spin but opposite parity ---
without loss of generality we can consider baryons of spin and
isospin 1/2 --- and expand the matrix elements $j^{a}_{5\, \mu}$
in terms of form factors,
\begin{equation}
\label{current}
\langle B_{+} |j_{5\ \mu}^{a}|B_{-} \rangle = \bar
u(p,s)\left(\gamma_{\mu}g_{A}(q^{2})+q_{\mu}g_{P}(q^{2})
+i\sigma_{\mu\nu}q^{\nu}g_{M}(q^{2}) \right)t^{a}\,u(p',s')
\end{equation}
where $q^{\mu}=p^{\mu}-p'^{\mu}$.  Conservation of the axial
current demands that $q^{\mu}\langle B_{+}|j_{5\
\mu}^{a}|B_{-}\rangle =0$, which gives
\begin{equation}
\label{gt}
 \Delta m g_{A}(\Delta m^{2})+\Delta m^{2}g_{P}(\Delta m^{2})=0
\end{equation}
for states at rest ($\vec q=0$), where $\Delta m=m_{+}-m_{-}$.
$g_{P}(q^{2})$ has a pion pole at $q^{2}=0$ with residue
$f_{\pi}g_{\pi B_{+}B_{-}}$,  where $g_{\pi B_{+}B_{-}}$ is  the
$s$-wave $B_{+}B_{-}\pi$ coupling constant. Then for small $\Delta
m$,
\begin{equation}
\Delta m g_{A}+f_{\pi}g_{\pi B_{+}B_{-}}+{\cal O}(\Delta m^{2})=0
\label{fullgt}
\end{equation}
where $g_{A}=g_{A}(0)$.

Suppose for some reason that $B_{+}$ and $B_{-}$ were degenerate.
Then $\Delta m=0$ and eq.~(\ref{fullgt}) requires $g_{\pi
B_{+}B_{-}}=0$ --- the pion must decouple at zero momentum.  The
argument goes through for states of different isospin with $\Delta
I=1$, as well as for mesons and strange and heavy hadrons. Thus
any states of the same spin, isospin, {\it etc.\/}, but opposite
parity that have the same mass must decouple from pions at zero
momentum.  This is a consequence of axial current conservation. So
chiral symmetry does not guarantee the states to be degenerate,
but if they are, it forces them to decouple from pions at zero
momentum, {\it i.e.\/} in the $s$-wave.

The thesis of
Refs.~\cite{Jido:1999hd,Jido:2001nt,glozman1,glozman2,review} is
that hadrons show parity doubling because  at masses of order 1.5
--- 3 GeV (where the data show evidence of parity doubling) chiral
symmetry is approximately restored in the hadron spectrum.  By
this they mean that the symmetry is realized (approximately)  in
Wigner-Weyl mode with the appearance of (approximately) degenerate
multiplets that transform into one another linearly under chiral
rotations.

The irreducible representations of $SU(2)_{L}\times SU(2)_{R}$ are
labelled by the Casimirs of the left/right-handed group factors
$(I_L, I_R)$ which are related to the usual isospin and its axial
partner by $I_{L,R}^a = T^a \mp X^a$. The representation
$(I_L,I_R)$ contains $(2I_L +1 )(2I_R +1)$ states, with total
isospin taking all possible values compatible with $|I_L - I_R|
\leq I \leq I_L + I_R$. Parity exchanges the left and right-handed
isospins $I_L \leftrightarrow I_R$,  such that these irreducible
representations are in general not parity invariant, unless $I_L =
I_R$.  We are interested in the baryon states such as $N, \Delta$
with isospins $1/2$ and $3/2$. These states can be contained only
in chiral multiplets, $(1/2,0)$, $(1/2,1)$, $(3/2,0)$, $(3/2,1)$,
$\ldots$, which are not parity invariant. It is easy to form
parity eigenstates by taking sums and differences of mirror chiral
representations
\begin{eqnarray} \label{mirror}
H_+ \sim (I_L, I_R) \oplus (I_R, I_L)\,,\qquad
 H_- \sim (I_L, I_R) \ominus (I_R, I_L)
\end{eqnarray}
In this example, $H_\pm$ represents  parity-even(odd) hadrons.
Therefore a Wigner-Weyl representation of $SU(2)_{L}\times
SU(2)_{R}$ generically includes degnerate multiplets of both
positive and negative parity and  a range of isospins.

In addition to degeneracy, the linear realization of the
$SU(2)_{L}\times SU(2)_{R}$ predicts relations between axial and
vector charges typical of a non-Abelian symmetry.  In
Ref.~\cite{us} we showed that decoupling of pions implied by
eq.~(\ref{gt}) in the parity doubling limit is not a curiosity.
Instead hadrons can only form Wigner-Weyl representations of
$SU(2)_{L}\times SU(2)_{R}$ if they \emph{completely decouple from
pions}.  Indeed this is just the mechanism proposed in
Refs.~\cite{glozman1,glozman2,review}, where it is argued that
states high in the spectrum are generically insensitive to the
effects of spontaneous symmetry breaking. Whether or not one
accepts these arguments, this dynamical picture makes many
predictions that can be tested.  First and foremost is the
prediction that the parity doubled states should decouple from
pions, not only in the $s$-wave at zero momentum as required by
the Goldberger-Treiman relation, eq.~(\ref{gt}), but in higher
partial waves as well.  Also, the arguments of
Ref.~\cite{glozman1,glozman2,review} in the parity doubled domain
of the spectrum apply whether or not the particular states are
degenerate.  All the pion transition matrix elements should be
small to the extent that chiral symmetry is ``restored''.  Second,
larger multiplets including states of different isospin should be
found in the spectrum.  Glozman and
Klempt\cite{Glozman:2004gk,Klempt:2002tt} have made some initial
attempts to classify states into such multiplets, with varying
amounts of success.  Third, strange and heavy hadrons should show
the same phenomenon --- the emergence of explicit $SU(2)_{L}\times
SU(2)_{R}$ symmetry high in the spectrum --- since the heavy
quarks are merely spectators to this dynamical mechanism. Finally,
the phenomenon should be generic.  If some well established states
high in the spectrum \emph{cannot} be classified into
representations of $SU(2)_{L}\times SU(2)_{R}$, it is evidence
that other dynamics is at work.

For the sake of completeness, we review the argument of
Ref.~\cite{us} that pions must decouple from states that exhibit
the full structure of $SU(2)_{L}\times SU(2)_{R}$ symmetry.
Parity invariance of QCD requires that physical states are parity
eigenstates. As explained above, this can be realized only if the
physical states transform according to sums and differences of
chiral representations as in eq.~(\ref{mirror}). Thus the combined
consequence of chiral symmetry realized in WW mode and parity
invariance of QCD, is the appearance of parity doubling in the
hadron spectrum.

However, we know that in Nature chiral symmetry is realized in
Nambu-Goldstone mode with the appearance of Goldstone bosons.
Actually two distinct possible NG realizations have been discussed
in the literature, distinguished by the invariance of the vacuum.
In the usual realization $SU(2)_L \times SU(2)_R$ breaks to
isospin. Long ago, Dashen pointed out that $SU(n_f)_L \times
SU(n_f)_R$   could break to $SU(n_f)_V \times Z^{A}_{n_f}$, where
$Z^{A}_{n_{f}}$ is the center of axial $SU(n_{f})$\cite{dashen}.
We consider the usual scenario first, and return to Dashen's
alternative at the end of this subsection.

Consider for definiteness two spin-1/2 baryons of positive $(B_+)$
and negative $(B_-)$ parity, and unspecified isospin
$I$\footnote{The example which follows is equivalent to the
``mirror model'' presented in Ref.~\cite{DeTar:1988kn} in the
context of the linear sigma model.}. Assume that they transform
linearly under chiral symmetry
\begin{eqnarray}
\label{linear}
&& [T^a, B_+^i] = - t^a_{ij} B_+^j\,, \qquad [T^a, B_-^i] = - t^a_{ij}B_-^{j} \\
&& [X^a, B_+^i] = t^a_{ij} B_-^j\,, \qquad \,\,\,\, [X^a, B_-^i] =
t^a_{ij}B_+^{j}
\end{eqnarray}
with $i,j$ isospin indices. This corresponds to the
$SU(2)_{L}\times SU(2)_{R}$ representations $B_+ \sim (I, 0) + (0,
I)$ and $B_- \sim (I,0) - (0, I)$. The most general chirally and
parity invariant effective  Lagrangian \emph{containing only}
$B_\pm$, reads
\begin{eqnarray}\label{L0}
{\cal L} = \bar B_+ i\partslash B_+ + \bar B_- i\partslash B_- -
m_0 (\bar B_+ B_+ + \bar B_- B_-) + \cdots
\end{eqnarray}
where the ellipses denote terms of higher order in the chiral
expansion. The two hadrons are degenerate, in agreement with the
arguments of \cite{review}. In addition, the matrix elements of
the axial current are also fixed by the symmetry, via the Noether
theorem.

So far we have neglected the possible coupling of $B_{+}$ and
$B_{-}$ to pions.   The pion field transforms nontrivially under
the chiral group.  This has the consequence that additional
operators, constructed from the baryon fields $B_\pm$ and $\pi^a$,
can be added to the effective Lagrangian, eq.~(\ref{L0}). For
example, using Weinberg's choice\cite{Weinberg:1968de,book} for
the pion non-linear transformation under axial rotations,
\begin{equation}
[X^a, \pi^b] = - if^{{ab}}(\pi) =
-i\left(\delta^{ab}\frac{1}{2}(1-\vec \pi ^{2}) + \pi^a \pi^b
\right)\ ,
\label{pionrule}
\end{equation}
(where we expressed the pion field in units of $f_\pi = 131 $ MeV)
another dimension three, chirally invariant operator can be added
to eq.~(\ref{L0}),
\begin{equation}
\label{Llin}
 \delta{\cal L}
 = m_1  \Big( \bar B\frac{1- \pi^2}{1+\pi^2}B
 - \bar B \frac{4i\pi^a t^a}{1+\pi^2} B'- (B\leftrightarrow B') \Big)\nonumber
\end{equation}
Rediagonalizing the terms quadratic in $B_{\pm}$ we see that the
new term has the effect of splitting the degeneracy:  the new mass
eigenvalues are $m_{0}\pm m_{1}$.  $m_{1}$ is the coupling
constant set to zero by the generalized Goldberger-Treiman
relation if the hadrons $B_{+}$ and $B_{-}$ are to be degenerate.

The effects of pion couplings are most  conveniently displayed by
redefining the baryon fields,
\begin{eqnarray}\label{lin2nl}
\tilde B_+ = \frac{B_+ - 2  i \pi^a t^a B_-}
   {\sqrt{1+ \pi^2}} \,, \qquad
\tilde B_- = \frac{B_- - 2 i \pi^a t^a B_+}
   {\sqrt{1+  \pi^2}}
\end{eqnarray}
The new fields have also positive $(\tilde B_+)$ and negative
$(\tilde B_-)$ parity, but transform nonlinearly under chiral
rotations. Specifically,
\begin{eqnarray}
\label{W}
[X^a , \tilde B_{\pm \,i}] = v_0(\pi^2) \varepsilon_{abc} \pi^c
t^b_{ij} \tilde B_{\pm \,j}\,,
\end{eqnarray}
with $v_0(\pi^2)$ a function which depends on the particular
definition of the pion field. We use here and in the following the
definition adopted in Ref.~\cite{book}, which corresponds to
$v_0(x)=1$. Most importantly, the new states $\tilde B_\pm$ do not
transform into one another under chiral transformations.  Instead
of taking $B_+ \leftrightarrow B_-$ as in eqs.~(\ref{linear}), the
axial chiral rotation of the fields defined in  eq.~(\ref{lin2nl})
transforms each of $\tilde B_i$ into itself, plus a number of
pions. This agrees with the intuitive idea that an axial rotation
does not transform  hadrons  within a  chiral  multiplet, but just
rotates their isospins and creates pions from vacuum.

It is easy  to construct the most general chirally invariant
effective Lagrangian describing the dynamics of the baryons
$\tilde B_\pm$ following the usual rules of chiral effective field
theory \cite{Cheff,Weinberg:1968de,book}. When rewritten in terms
of the original fields, $B_\pm$, we discover several other
possible chirally invariant terms, allowed when pions transform
nonlinearly, but omitted from eq.~(\ref{L0}). They take the
simplest form when expressed in terms of the $\tilde B_\pm$
fields, and are given by
\begin{eqnarray}\label{Lnl}
{\cal L} &=& \bar {\tilde B}_+ (i\partslash -\varepsilon^{abc}
\pi^a \Dslash \pi^b t^c)  \tilde B_+   -
\bar {\tilde B}_+ (\Dslash \pi^a) t^a \tilde B_- \nonumber \\
& & \hspace{-1cm} +(\tilde B_+ \leftrightarrow \tilde B_-) - (m_0
- m_1) \bar {\tilde B}_+ \tilde B_+ - (m_0 + m_1) \bar {\tilde
B}_- \tilde B_- + \delta {\cal L}_2
\end{eqnarray}
where $D_\mu \pi^a = 2\partial_\mu \pi^a/(1+\pi^2)$ is the
covariant derivative of the pion field. The additional terms
$\delta {\cal L}_2$ contain three operators which couple the pions
derivatively to the baryons
\begin{eqnarray}
\delta {\cal L}_{2}&=&c_{2}[  \bar {\tilde B}_+ (\Dslash
\pi^a)\gamma_5 t^a \tilde B_+
              + \bar {\tilde B}_- (\Dslash \pi^a)\gamma_5 t^a \tilde B_-]\nonumber\\
&+&c_{3}[ \bar {\tilde B}_+ (\Dslash \pi^a) t^a \tilde B_-
              + \bar B_- (\Dslash \pi^a) t^a \tilde B_+ ]\nonumber \\
&+&c_{4}[\bar {\tilde B}_+ (\Dslash \pi^a)\gamma_5 t^a  \tilde B_+
- \bar {\tilde B}_- (\Dslash \pi^a)\gamma_5 t^a \tilde B_-]
\end{eqnarray}
Note that the baryons $\tilde B_\pm$ are not degenerate, but are
split by $2m_1$. Similar results are obtained for the matrix
elements of the axial current, which is obtained from the Noether
theorem
\begin{eqnarray}\label{Noether}
A_\mu^a &=& (c_2 + c_4) \bar {\tilde B}_+ \gamma_\mu \gamma_5 t^a
\tilde B_+ +
(c_2 - c_4) \bar {\tilde B}_- \gamma_\mu \gamma_5 t^a {\tilde B}_- \nonumber \\
&+& (1-  c_3) (\bar {\tilde B}_+ \gamma_\mu t^a \tilde B_- +
\mbox{h.c.}) + \mbox{(pion terms)}
\end{eqnarray}
No predictions are obtained from chiral symmetry for these
couplings, which can take any values.

The scenario of ``effective chiral symmetry restoration''
discussed by the authors of Ref.~\cite{review} requires that the
coefficients of the chirally invariant operators $m_1, c_{2-4}$ in
this effective Lagrangian be suppressed. The alternative
explanations for parity doubling described in the following
subsections, would, in the language of $SU(2)_{L}\times SU(2)_{R}$
symmetry restoration, amount to a dynamical   mechanism for the
suppression of $m_{1}$ and $c_{2-4}$. For further discussion of
parity doubling and $SU(2)_{L}\times SU(2)_{R}$ symmetry
restoration, see Ref.~\cite{us}. Note that these results are
specific to the assignment of $B_\pm$ to the chiral
representations $(0, \frac12), (\frac12, 0)$, and a different
chiral assignment will give different predictions for the pion
couplings.

Finally, for completeness, we describe the alternative scenario
suggested in 1969 by Dashen \cite{dashen}, and revisited recently
by Kogan, Kovner, and Shifman\cite{kks}, wherein the vacuum is
left invariant not only by the vector generators $\{T^a\}$, but
also by a discrete symmetry  $Z^{A}_{n_f}$  (the center symmetry
of $SU(n_{f})_A$). This scenario has no implications for parity
doubling unless $n_{f}\ge 3$, so we restrict our consideration to
the minimal case $n_f=3$. Although at least three light flavors
are required, this scenario would have nontrivial implications for
the nonstrange baryons, which are our main focus here.

\begin{figure}[h]
\label{dashenSR}
\includegraphics[width=7cm]{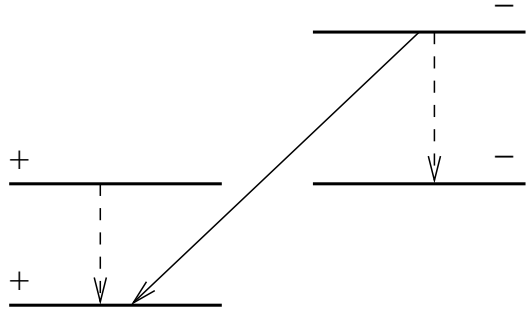}
\caption{\small Typical baryon mass spectrum in the Dashen
scenario for chiral symmetry breaking. Shown are two singlets of
positive and negative parity, and a parity doublet. Pion
transitions $h_1 \to h_2 \pi$ between singlets and doublets are
forbidden by the symmetry (dashed arrows), while transitions among
singlets and among doublets are allowed (solid arrow).}
\end{figure}

The center of axial $SU(3)$ contains the operator $Z =
\exp(\frac{2\pi i}{3}X_{8} )$ along with  $Z^2$ and the identity.
Since $Z$ does not commute with parity, the complete discrete
symmetry of the vacuum is the dihedral group $D_3 = (1, Z, Z^2, P,
PZ, PZ^2)$. According to Coleman's theorem \cite{coleman}, the
same symmetry should also be apparent in the hadron spectrum.
Since $D_3$ has both one- and two-dimensional irreducible
representations, the Dashen scenario of chiral symmetry
realization could give rise to parity doubling in the hadron
spectrum \footnote{This does not occur in the two flavor case
because $Z_{2}\times$ parity is Abelian and has only one
dimensional irreducible representations.}. More precisely, there
are one-dimensional representations of either parity, and
two-dimensional representations, containing parity doublets.

Since mesons are singlets under the discrete $D_3$ group, this
symmetry gives a selection rule for the pion strong decays between
hadrons:  the $h^* \to h\pi$ decay is allowed only between
singlets, or only between doublets, but not between singlets and
doublets as illustrated in Fig.~\ref{dashenSR}. No clear evidence
for this selection rule is observed among the baryons --- in fact
Dashen points out several problems with it that still pertain
today --- although symmetry breaking effects through mass terms
could obscure it. We refer to the original work \cite{dashen} for
details.

In the Dashen scenario the  $\langle \bar q q\rangle$  condensate
vanishes in the massless quark limit,  and $f_{\pi}\sim m_{q}$ as
well. Phenomenological analysis of the pattern of chiral symmetry
breaking~\cite{kks,knecht} does not rule out this behavior.
However Kogan, Kovner, and Shifman have shown that it is in
conflict with rigorous QCD inequalities~\cite{inequalities}.  So
we do not consider the Dashen scenario further as a possible
origin for parity doubling.

\subsection{ Dynamical Suppression of $\mathbf{U(1)_{A}}$ Symmetry Violation}

In this section we adopt the standard picture of $U(1)_{A}$
symmetry violation in QCD as summarized, for example, by 't
Hooft\cite{thooft}.  The $U(1)_{A}$ symmetry is broken in the
quantum Hamiltonian by quark masses and by the triangle anomaly.
Since we are interested in baryons made of $u$, $d$, and $s$
quarks, we consider QCD with three flavors.  On account of the
explicit symmetry breaking, the flavor singlet axial current,
$j^{\,\mu}_{5}$, has a divergence, and its charge, $Q_{5}(t) =
\int d^3 x j^{\,0}_{5}(t,\vec x)$, does not commute with the
Hamiltonian,
\begin{equation}
\label{abj}
[H, Q_{5}] = i\frac{d}{dt}Q_5 = i\int d^3 x \p_{\mu}j^{\,\mu}_{5}
\,,\qquad \p_{\mu}j^{\,\mu}_{5}= \sum_{q=u,d,s}2m_{q}i\bar
q\gamma_{5} q -i\frac{3g^{2}}{16\pi^{2}}{\rm Tr}\tilde
F^{\mu\nu}F_{\mu\nu}
\end{equation}
Because there are no massless particles coupled to
$j^{\,\mu}_{5}$, the singlet axial charge is a well defined
operator.  This contrasts with the isospin axial charges,
$Q_{5}^{a},\, a=1,2,3$, which become ill-defined in the chiral
limit.

Some of the consequences of $U(1)_{A}$ invariance, including
parity doubling, can be obtained if the matrix elements of
$[H,Q_{5}]$ are suppressed in a subspace of the space of states
where the spectrum is sparse.  The operator $Q_{5}$ can create
$\eta'$ mesons from the vacuum, so it clearly contains frequency
components of order 1 GeV.  We believe this obstructs restoration
of $U(1)_{A}$ symmetry in a \emph{strong}, operator, sense.
However a \emph{weak} form, applied only to matrix elements within
the subspace, seems possible and could lead to parity doubling.

\begin{table}
\caption{Comparison of the physical manifestations of the scenario
discussed here of ``suppression of $U(1)_A$ symmetry breaking
(SB)'', with those of ``$U(1)_A$ restoration'', which would be
obtained by attempting to realize $U(1)_A$ in Wigner-Weyl mode. }
\begin{tabular}{|c|c|c|}
\hline
Prediction & $U(1)_A$ restoration & Suppression of $U(1)_A$ SB \\
\hline
$Q_5 |0\> = 0$ & yes & no \\
parity doubling & yes & yes \\
mass states are $Q_5$ eigenstates & yes & no \\
\hline
\end{tabular}
\label{table2}
\end{table}

Suppose first, for simplicity, that there were a subspace,
${\mathfrak h}$, of the Hilbert space of QCD, in which the matrix
elements of $\p_{\mu}j^{\,\mu}_{5}$ were exactly zero. One
possibility is that $Q_{5}$ also has no non-zero matrix elements
in this subspace.  In that case the symmetry has no implications
for the spectrum in ${\mathfrak h}$.  Alternatively, suppose
$Q_{5}$ has a non-zero matrix element between two mass eigenstates
$|B_{+}\>$ and $|B_{-}\>$ in ${\mathfrak h}$. $B_{+}$ and $B_{-}$
must have the same flavor and angular momentum quantum numbers but
opposite parity, denoted by the $\pm$ sign.  According to
eq.~(\ref{abj}) they must also have the same mass if
$\p_{\mu}j^{\,\mu}_{5}$ can be neglected as we have assumed,
\begin{equation}
\<B_{+}|\left[Q_{5},H\right]|B_{-}\> =
\left(m(B_{-})-m(B_{+})\right)\<B_{+}| Q_{5}|B_{-}\> =0
\label{conserve}
\end{equation}
If $\<B_{+}|Q_{5}|B_{-}\> \ne 0$, as we have assumed, then
$m(B_{+})=m(B_{-})$. So the spectrum of states with a given $J$,
$I$, and $S$, consists either of isolated states which have no
non-zero matrix elements of $Q_{5}$ within ${\mathfrak h}$, or of
degenerate families of both positive and negative parity connected
by $Q_{5}$.  Excluding accidental degeneracies, the result is
parity doubling.

We do not believe that stronger conditions can be imposed:
\begin{enumerate}
\item[a.]  $[H,Q_{5}]=0$  and $Q_5 |0\rangle =0$

This condition is not consistent with the QCD lagrangian.
Phenomenologically, it leads to parity doubling throughout the
baryon spectrum. Since parity doubling is at best an approximate
symmetry in  a subspace, and clearly does not hold for the
lightest states, this condition is not tenable.

\item[b.] $[H,Q_{5}]|B\>=0$ if $B\in {\mathfrak h}$

It is easy to show that this condition leads to parity doubling of
the state $|B\>$, however we believe it is too strong.  In
particular, it implies that the state $Q_{5}|B\>$ must be an
eigenstate of $H$ with the same mass as $|B\>$. We expect that the
operator $Q_{5}$ generically  creates $\eta'$ mesons, so the
amplitude $\<B'\eta'|Q_{5}|B\>$, where $|B'\>$ might be the
original state $|B\>$ or some other baryon, should not be small.
The weak condition we have adopted does not preclude the
possibility that the axial charge acting on one member of a parity
doublet has a significant amplitude to produce an $\eta'$.
\end{enumerate}

If the matrix elements of $\p_{\mu}j^{\,\mu}_{5}$ are small, but
not zero, in ${\mathfrak h}$, then the degeneracies are no longer
exact. It is interesting to consider this case from the point of
view of the $U(1)_{A}$ Goldberger-Treiman relation and to contrast
it with the $SU(2)_{L}\times SU(2)_{R}$ case.\footnote{The
$U(1)_{A}$ Goldberger-Treiman relation was considered for nucleons
in \cite{shoreandveneziano}.} Again taking $B_{\pm}$ to be
spin-$1/2$ baryons, the matrix element of the flavor singlet axial
current and its divergence  can be parameterized in terms of
invariant form factors,
\begin{eqnarray}
\label{gtr}
\<B_{+}|j_{5 \, \mu}(0)|B_{-}\>&=&\bar u(p_{+})\left(
\gamma_{\mu}g_{A}(q^{2})+q_{\mu}g_{P}(q^{2})+i\sigma_{\mu\nu}q^{\nu}g_{M}(q^{2})\right)u(p_{-})\nonumber\\
\<B_{+}|  \p^{\mu}j_{5\, \mu} |B_{-}\>&=&\bar u(p_{+})d_{A}(q^{2})
u(p_{-})
\end{eqnarray}
Taking the states to be at rest, $q^{2}=(q^{0})^{2}=\Delta m^{2}$.
Then contracting $q^{\mu}$ with the first of eqs.~(\ref{gtr}) we
obtain
\begin{equation}
\label{gtr2}
\Delta m \left(g_{A}(\Delta m^{2}) + \Delta m  g_{P}(\Delta
m^{2})\right) = d_{ A}(\Delta m^{2})
\end{equation}

Note that the quantity in parentheses is the matrix element of the
axial charge: $\<B_{+}| Q_{5}|B_{-}\>\propto g_{A}(\Delta
m^{2})+\Delta mg_{P}(\Delta m^{2})$. The flavor singlet case
differs   from the flavor non-singlet case in that there is no
zero mass pole in the induced pseudoscalar form factor, $g_{P}$.
Indeed, the nearest significant singularity in the pseudoscalar
channel is the $\eta'$ at nearly 1 GeV. Therefore, if the matrix
elements of $\p_{\mu}j^{\,\mu}_{5}$ are small, the induced
pseudoscalar contribution is negligible (second order in $\Delta
m^{2}$), leaving,
\begin{equation}
\label{uasymm}
\Delta m   g_{A}(0)\approx d_{A}(0)
\end{equation}

Eq.~(\ref{uasymm}) allows us to formulate the problem of
``suppression of $U(1)_{A}$ symmetry violation'': Are there states
for which $d_{A}(0)$ is small and $g_{A}(0)$ is not?  While we
cannot answer the question decisively,  it is well posed. At least
in principle, this question can be answered using lattice QCD
computations of the form factor $d_A(q^2)$. In the chiral limit,
this coincides with the form factor of the gluonic term in the
anomaly.

We emphasize that the scenario of ``suppression of $U(1)_{A}$
symmetry violation'' is very different from that of $U(1)_A$
restoration, or $U(1)_A$ realized in Wigner-Weyl mode. For
convenience we summarize in Table \ref{table2} the similarities
and differences between these two scenarios. In particular, we do
not require the invariance of the vacuum under $U(1)_A$
transformations.

Quark models suggest that the matrix elements of the singlet axial
current between states with similar quark configurations but
opposite parity are of order unity. For example the singlet axial
current in the bag model changes a quark with angular momentum $j$
and positive parity into a quark with angular momentum $j$ and
negative parity with amplitude of order unity.  For such states
$\Delta m\sim d_{A}(0)$, and suppression of the matrix elements of
the divergence of the singlet axial current would result in
approximate parity doubling. Other states, for which $g_{A}(0)$ is
smaller, need not be nearly degenerate, even though $d_{A}(0)$ may
be small.

The divergence of the flavor singlet axial current gets
contributions from the anomaly and from quark masses. Is it
possible that the matrix elements of $\tilde FF$ and $m_{s}\bar
s\gamma_{5}s$ are unusually small at moderate excitations in the
baryon spectrum? Perhaps instanton fluctuations are suppressed at
high excitation in the baryon spectrum as they are at short
distances.  We are not aware of any work along this
direction.\footnote{In Ref.~\cite{Kochelev:2005vd} Shifman's
analysis\cite{shifman} is extended to flavor singlet correlators,
where instanton contributions determine the convergence of sum
rules analogous to Weinberg's\cite{WeinbergSR}. Our analysis
suggests that the proper quantity to analyze would be matrix
elements of $\tilde FF$ between baryon states of identical $J$,
$I$, and $S$, but opposite parity, possibly using instanton
methods along the lines of ref.~\cite{inst}.} We do not know how
to estimate the matrix elements of $F\tilde F$ between excited
hadrons, and leave it for further study.

The generalization of suppression of $U(1)_{A}$ symmetry violation
to strange hadrons is not straight forward.  In the limit
$m_{s}=0$ $SU(3)_{V}$ becomes exact and therefore suppression of
$\langle F\tilde F\rangle$ would lead to parity doubling among
strange as well as non-strange baryons in octets and decuplets.
When the strange quark mass is increased, the flavor singlet
current has an increasing divergence proportional to $m_{s}\bar
s\gamma_{5}s$.  However there is a linear combination of flavor
singlet and non-singlet currents that includes only $u$ and $d$
quarks, $\bar J_{5}^{\,\mu}\equiv\bar
u\gamma^{\mu}\gamma_{5}u+\bar d\gamma^{\mu}\gamma_{5}d$, whose
divergence comes only from the axial anomaly,
$$
\partial_{\mu}\bar J^{\,\mu}_{5} =  -i\frac{2g^{2}}{16\pi^{2}}{\rm Tr}\tilde
F^{\mu\nu}F_{\mu\nu}
$$
The discussion of parity doubling among non-strange hadrons can be
repeated without modification using $\bar J^{\,\mu}$ instead of
the flavor singlet current, $j^{\,\mu}_{5}$.  However $\langle
N^{\pm},\Delta^{\pm} |F\tilde F|N^{\mp},\Delta^{\mp}\rangle
\approx 0$ does not imply that $F\tilde F\approx 0$ for their
strange $SU(3)$ partners.  For example, one might model the
effects of the anomaly by the 't Hooft determinant, ${\rm det\
}\bar q_{R}q_{L}+{\rm det\ } \bar q_{L}q_{R}$, which is a
six-quark operator of the form $\bar u\bar d\bar s
uds$\cite{gerard}.  This operator has tree level matrix elements
in $\Lambda$'s and $\Sigma$'s, but only contracted loop
contributions in non-strange baryons.  Without a better model of
hadron matrix elements of $F\tilde F$, it is not possible to
conclude that parity doubling among non-strange baryons implies
doubling among the analogous strange states.

Finally it is worth emphasizing that this sort of symmetry
restoration occurs frequently in quantum systems.  Since the
phenomenon involves \emph{explicit} rather than \emph{spontaneous}
symmetry violation, we can find examples in ordinary quantum
mechanics.  Consider, for example, an atom in an external electric
quadrupole field, $\delta H_{SB}\propto
e(r_{i}r_{j}-\fracs13\delta_{ij})Q^{ij}$.  Although rotational
invariance is broken in general, it is unbroken among $j=1/2$
states where $\langle \psi_{\half,m}| [\delta H_{SB},\vec
J]|\psi'_{\half,m'}\rangle=0$ even though $[\delta H_{SB},\vec
J]|\psi_{\half,m}\rangle\ne 0$ in general.

\subsection{Parity Doubling and Chiral Symmetry Restoration at Short Distances}

The idea that $SU(2)_L \times SU(2)_R$ chiral symmetry is restored
at asymptotically large Euclidean momenta (short distances) in
current correlators in QCD dates back to Weinberg's 1967 paper on
spectral function sum rules. These sum rules relate moments of the
spectral functions of vector and axial currents. However, turning
them into relations among individual meson states is a different
problem, and requires additional theoretical input.

 Recently, such relations have been discussed by several authors
~\cite{shifman,silas}, who argued that parity doubling might be a
consequence of the chiral invariance of current correlators at
short distances in QCD and other dynamical assumptions. Usually
these arguments are framed  in terms of ``chiral symmetry
restoration'', so it is interesting to ask, in the spirit of
Ref.~\cite{us}, what additional dynamics are responsible for
 suppressing the chirally invariant operators necessary to obtain a Wigner-Weyl
realization of chiral symmetry in the spectrum.

Refs.~\cite{shifman,silas}  discuss mesons at high excitation and
large $N_{c}$. The arguments do not go over to the baryon sector
unchanged~\cite{shifmanprivate} since baryons have different large
$N_{c}$ systematics than mesons, and baryon correlators have
different short distance behavior than meson current correlators.
Furthermore our phenomenological analysis pertains to hadrons in
the 1.0 --- 2.5 GeV region, and includes many resonances that are
the lightest states in a particular isospin and angular momentum
channel, where such an analysis certainly would not apply.
Nevertheless it remains interesting to examine the assumptions
required to obtain parity doubling in this simpler limiting case.
Ref.~\cite{silas} discusses relations among vector and axial
mesons, while Ref.~\cite{shifman} presents similar relations for
pseudoscalar and scalar mesons. For simplicity, we phrase our
discussion in terms of the latter case.

Shifman compares the correlators of two scalar currents,
$\Pi(Q^{2})$, with the correlator of two pseudoscalar currents,
$\tilde\Pi(Q^{2})$, in the Euclidean domain, $Q^{2}= -q^{2}>0$.
Shifman assumes that the difference, $\Delta\Pi(Q^{2})\equiv
\Pi(Q^{2})-\tilde\Pi(Q^{2})$ obeys an unsubtracted dispersion
relation,
\begin{equation}
\label{disp}
\Delta\Pi(Q^{2})\equiv \Pi(Q^{2})-\tilde\Pi(Q^{2}) = - \int
d\sigma^{2}\frac{\Delta\rho(\sigma^{2})}{Q^{2}+\sigma^{2}}
\end{equation}
where the spectral functions $\rho(\sigma^{2})$ and $\tilde
\rho(\sigma^{2})$ (with $\Delta\rho\equiv \rho-\tilde\rho$)
receive contributions from physical intermediate states of mass
$\sigma$ created by the currents from the vacuum.  Well
established arguments based on the operator product expansion and
QCD sum rules\cite{swz} require that $\Delta\Pi(Q^{2})$ vanishes
like $1/Q^{4}$ as $Q^{2}\to\infty$,
 \begin{equation}
 \label{correlators}
 \Delta\Pi(Q^{2})\underset{Q^{2}\to\infty}{\sim}  \frac{\<\overline\Psi\Psi\>}{Q^{4}}^{2}
  \end{equation}
where $\<\overline\Psi\Psi\>$ is the chiral symmetry violating
condensate. Each of the correlators, $\Pi$ and $\tilde\Pi$,
individually go like $Q^{2}\log Q^{2}$ as $Q^{2}\to\infty$, so the
fact that their difference vanishes as $1/Q^{4}$ is evidence of
the rapid restoration of chiral symmetry \emph{at short distances}
in QCD.  Combined with the unsubtracted dispersion relation, the
condition that there is no term $\sim 1/Q^{2}$ in
$\Delta\Pi(Q^{2})$ requires a superconvergence relation,
\begin{equation}
\label{super}
\int d\sigma^{2}\Delta\rho(\sigma^{2})=0
\end{equation}

Eq.~(\ref{super}) is the \emph{only} input from chiral symmetry in
the argument of Ref.~\cite{shifman}. Convergence of the $\sigma$
integration requires that the spectral functions, each of which
grows like $\sigma^{2}$  must quickly become equal at large
$\sigma$. Naively one would think that this constraint applies at
asymptotically large masses where broad, overlapping resonances
build up continua in $\rho$ and $\tilde\rho$.  Strong additional
assumptions are needed to make inferences about parity doubling.
The principal assumption is that QCD may be described by a limit
in which the number of colors, $N_{c}$, is taken to infinity
\emph{before} considering high radial excitations (labeled by a
radial quantum number $n$).

 To explore the implications of this
unusual order of limits, we analyze the problem at large but fixed
$N_{c}$ and extract the leading order predictions at the end.
First, let us recap the argument of Ref.~\cite{shifman}, taking
$N_{c}\to\infty$ first.   Then, the spectral function $\rho$
($\tilde\rho$) receives contributions only from \emph{zero width}
scalar (pseudoscalar) meson resonances, with masses $M_{n}$
($\tilde M_{n}$)  and coupling to the current $f_{n}$ ($\tilde
f_{n}$).  Continuum contributions are suppressed by powers of
$N_{c}$. For reference, the squared masses, squared widths, and
couplings of the mesons are assumed to have the following
dependence on $n$ and $N_{c}$
\begin{eqnarray}
\label{asympt}
 M_{n}^{2}, \tilde M_{n}^{2}&\sim& \Lambda^{2}n\nonumber\\
\Gamma_{n}^{2}, \tilde \Gamma_{n}^{2}&\sim&  \Lambda^{2}n/N_{c}^{2}\nonumber\\
f_{n}, \tilde f_{n} &\sim& N_{c}\Lambda^{4}n
\end{eqnarray}
where $\Lambda$ is the dynamical mass scale of QCD.  The $n$
dependence is motivated by string analogies,  the spectra of
linear potentials and the 't Hooft model\cite{shifman,thooft1dim}.
The factors of $N_{c}$ are in accord with standard large-$N_{c}$
counting arguments.

Ignoring the widths or equivalently, exchanging the limit
$N_{c}\to\infty$ with the sum over $n$, the spectral functions are
sums of poles on the real axis, so
\begin{equation}
\label{zerowidth}
\Delta\rho(\sigma^{2}) =
\sum_{n}\left(f_{n}\delta(\sigma^{2}-M_{n}^{2})-\tilde
f_{n}\delta(\sigma^{2}-\tilde M_{n}^{2})\right)
\end{equation}
and the superconvergence relation reads,
\begin{equation}
\label{superzero}
N_{c}\Lambda^{2}\sum_n\left(\alpha_{n}M_{n}^{2}-\tilde
\alpha_{n}\tilde M_{n}^{2}\right) = 0
\end{equation}
where we have defined $f_{n} =N_{c}\Lambda^{2}\alpha_{n}$, {\it
etc.\/}  This sum must converge at large $n$.  Taking
$\alpha_{n}=\tilde\alpha_{n}=\alpha$, independent of $n$ (an
assumption later relaxed in Ref.~\cite{shifman}) the weakest
behavior large $n$ behavior consistent with the convergence of the
sum in eq.~(\ref{superzero}) is
$$
M_{n}^{2}-\tilde M_{n}^{2}\underset{n\to\infty}{\sim}{\rm
alt}\frac{1}{n}
$$
where ``alt'' denotes an alternating sign.  Since $M_{n}\sim
\tilde M_{n}\sim \sqrt{n}$ at large $n$, Shifman finds $\Delta
M_{n}\equiv M_{n}-\tilde M_{n}\sim\ {\rm alt}\  n^{-3/2}$,
indicating that the spectrum achieves parity doubling rather
rapidly at large $n$\footnote{This derivation has been criticized
in Ref.~\cite{subtractions}, where it is pointed out that the
$N_{c}\to\infty$ correlator, $\Delta\Pi(Q^{2})$, does not satisfy
an unsubtracted dispersion relation, or equivalently that certain
interchanges of limits and sums/integrals, required by the
derivation, are not allowed by the asymptotic behavior of
$\Delta\rho(\sigma^{2})$ when $N_{c}$ is infinite.  By keeping
finite $N_{c}$ corrections (see the following paragraph), which
replace the $\delta$-functions in $\Delta\rho$ with finite width
resonances (in the manner proposed in Ref.~\cite{shifman2}), we
maintain the unsubtracted superconvergence relation and avoid any
problems with interchanges of limits.  By this approach we include
important physics, namely the fact that the resonant contributions
to the spectral functions merge into a continuum at large $n$ for
any finite $N_{c}$, and, happily, agree with the conclusions of
Ref.~\cite{subtractions} that asymptotic chiral symmetry and
conventional large $N_{c}$ arguments alone do not require parity
doubling.}.

For any fixed $N_{c}$ the widths of resonances grow at the same
rate as their masses (see eq.~(\ref{asympt})), so the spectral
functions approximate a continuum as $n\to\infty$ at fixed
$N_{c}$.  Let us keep track of these $1/N_{c}$ corrections.   The
effect of a finite width for the $n^{\rm th}$ meson on the
spectral functions was analyzed in Ref.~\cite{shifman2}.  When the
widths are finite the meson poles in the correlation functions
move into the lower half complex $q^{2}$-plane, and the
correlation functions develop cuts with branch points at the
thresholds for the decay of the $n^{\rm th}$ meson.  We refer to
Ref.~\cite{shifman2} for details.  It is straightforward to find
the modification of eq.~(\ref{superzero})
\begin{equation}
\label{superzero2}
N_{c}\Lambda^{2}\sum_n\left(\alpha_{n}(M_{n}^{2}-\frac{1}{\pi}\Gamma_{n}M_{n})-\tilde
\alpha_{n}(\tilde M_{n}^{2}-\frac{1}{\pi}\tilde\Gamma_{n}\tilde
M_{n})\right) = 0
\end{equation}
Repeating the same arguments that followed eq.~(\ref{superzero}),
we find
$$
\Delta
M_{n}-\frac{1}{\pi}\Delta\Gamma_{n}\underset{n\to\infty}{\sim}{\rm
alt} \ \frac{1}{n^{3/2}}
$$
where $\Delta \Gamma_{n}\equiv \Gamma_{n}-\tilde\Gamma_{n}$.
Referring back to eqs.~(\ref{asympt}) we see that
$\Gamma,\tilde\Gamma\sim \sqrt{n}/N_{c}$ to leading order in $n$
and $N_{c}$.  However we have no independent argument to fix the
large $n$ behavior of \emph{the difference} $\Delta\Gamma_{n}$. It
could, in principle go like a constant at large $n$, or like some
power of $n$ larger than $n^{-3/2}$.  If we parameterize our
ignorance by an exponent $\beta$, presumably smaller than $1/2$,
we can summarize the $1/N_{c}$ corrections to the analysis of
Ref.~\cite{shifman},
$$
\Delta M_{n}\underset{n\to\infty}\sim{\rm  alt}\ \frac{1}{n^{3/2}}
+C \frac{n^{\beta}}{N_{c}}
$$
where both $\beta$ and the constant $C$ are unknown.  Thus the
effect of $1/N_{c}$ (remember $N_{c}=3$ in our world) corrections
could completely destroy parity doubling. While further analysis
or more assumptions might indicate that these $1/N_{c}$ effects
can be neglected, this example serves well to illustrate the fact
that parity doubling does not follow from the chiral invariance of
QCD alone, nor from suppression of chiral symmetry
\emph{violating} effects.  In the example of Ref.~\cite{shifman}
the assumptions include those necessary to give the asymptotic
behavior summarized in eqs.~(\ref{asympt}) and to suppress the
effects of hadron widths at large but finite $N_{c}$.

\subsection{Parity Doubling and Intrinsic Deformation}

Parity doubling in the baryon spectrum might be a consequence of a
phenomenon well known in nuclear and molecular physics.  If a
system can be described by the collective quantization  of an
intrinsic state and a) the intrinsic state (spontaneously)
violates parity, and b) the deformation is relatively rigid, then
the low lying excitations of the system will display parity
doubling.  The phenomenon is well known, for example, in the
spectrum of ammonia.  Clear examples occur in nuclear physics as
well.  The existence of parity doubled states in nuclei like
$^{225}$Ra, with instrinsic octupole (``pear shaped'')
deformation, plays an important role in proposed sensitive
searches for $T$-violating electric dipole
moments\cite{Engel:2003rz,butler,cocks}.
%
Linear superpositions of the intrinsic state, $|\psi\>$ and its
parity image, $\hat P|\psi\>$ form parity eigenstates,
$|\psi_{\pm}\>=\frac{1}{\sqrt{2}}(|\psi\>\pm \hat P|\psi\>)$.  If
the deformation is rigid then the Hamiltonian matrix element
between $|\psi\>$ and $\hat P|\psi\>$ is small and the states of
opposite parity, but otherwise identical structure, are nearly
degenerate. This degeneracy survives quantization of the
collective coordinates that restores rotational symmetry resulting
in a tower of parity doubled excitations.

One can imagine dynamics that might lead to a similar phenomenon
in hadron spectroscopy \cite{iachello}, though a convincing
discussion is beyond our present understanding of hadrons in QCD.
If baryons on the leading Regge trajectory were well described as
a dumbbell-like structure with a quark on one end and a diquark on
the other\cite{wilczekselem}, then the intrinsic state would
violate reflection symmetry.   If tunneling of the odd quark from
one end to the other were suppressed enough, then parity doubling
would result.  If correct, this argument would predict parity
doubling for states of high angular momentum. There is some
evidence for this among nucleon ($I=1/2$) and $\Delta$ ($I=3/2$)
states, although the $\Delta$s are not very well known (see Table
\ref{table1}). Hyperons ($\Lambda$s and $\Sigma$s) are too poorly
known to provide much information.
\begin{table}
\caption{Parity doubled $I=1/2$ and $I=3/2$ baryons with angular
momentum $5/2$ or greater.  Data from the PDG\cite{PDG}.}
\begin{tabular}{|c|c|c|}
\hline
$B(J^{P})$ & PDG reliability & Mass (MeV)\\
\hline
$N(5/2^{-})$ & $****$ & 1675\\
$N(5/2^{+})$ & $****$ & 1680\\
\hline
$N(7/2^{+})$ & $**$ & 1990\\
$N(7/2^{-})$ & $****$ & 2190\\
\hline
$N(9/2^{+})$ & $****$ & 2220\\
$N(9/2^{-})$ & $****$ & 2250\\
\hline
\hline
$\Delta(5/2^{+})$ & $****$ & 1905\\
$\Delta(5/2^{-})$ & $***$ & 1930\\
\hline
$\Delta(7/2^{+})$ & $****$ & 1950\\
$\Delta(7/2^{-})$ & $*$ & 2200\\
\hline
$\Delta(9/2^{+})$ & $**$ & 2300\\
$\Delta(9/2^{-})$ & $**$ & 2400\\
\hline
\end{tabular}
\label{table1}
\end{table}
If this mechanism is responsible for parity doubling, one expects
the evidence for parity doubling should improve with increasing
angular momentum, and should be found in the spectrum of
$\Lambda$s and $\Sigma$s as well.  These predictions could be
tested in an experimental re-examination of baryon spectroscopy.
As a first attempt in this direction, we have repeated the
analysis of Section 2 on the non-strange baryons with the states
in Table \ref{table1} \emph{removed}.  The significance of the
signal for parity doubling remains virtually unchanged when the
high angular momentum states are removed.  Thus the phenomenon
cannot be entirely explained by collective coordinate quantization
of a parity-violating, enlongated intrinsic state.

\section{Implications from Mended Chiral Symmetry}

We explore in this Section some implications of chiral symmetry
using a comparatively less well-known approach. While not directly
relevant to the topic of parity doubling, this approach has
potentially interesting implications for the properties of baryon
resonances of opposite parity. As discussed in Section 2, chiral
symmetry realized in Nambu-Goldstone mode does not impose any
constraints on the masses and couplings of hadrons, beyond those
required by isospin. However, in a series of interesting papers,
Weinberg showed that the full chiral symmetry group may remain
manifest in a very special sense
\cite{Weinberg:1969hw,Weinberg:1990xn}. The symmetry is realized
on the matrix of the couplings of the Goldstone bosons, rather
than on the mass eigenstates, as typical for a symmetry realized
in Wigner-Weyl mode. Following Ref.~\cite{Weinberg:1990xn}, we
will refer to chiral symmetry realized in this fashion as ``mended
chiral symmetry''. When combined with the large energy asymptotics
of Goldstone boson scattering amplitudes following from Regge
theory,  the approach of
Refs.~\cite{Weinberg:1969hw,Weinberg:1990xn} gives also nontrivial
implications for the mass spectrum.

In this section we give a brief review of the approach of
Refs.~\cite{Weinberg:1969hw,Weinberg:1990xn}, and use it to
construct a chirally invariant model for a system of two nucleons
of opposite parity. In this model the two nucleons of opposite
parity are not necessarily degenerate, but their pion couplings
are exactly the same as predicted from a linear realization of
chiral symmetry. This approach provides such a natural realization
of such predictions of chiral symmetry ``restoration'' without
actual parity doubling.

The mended chiral symmetry approach is equally applicable to
mesons and baryons. In the original work
Ref.~\cite{Weinberg:1969hw} it was used to describe the quartet of
$J^P=0^\pm, 1^\pm$ mesons $(\sigma, \pi, \rho, a_1)$, and the
$(N,\Delta,N_1)$ system of the nucleon, delta and Roper baryons,
respectively. However, in the sector of mesons with zero helicity,
this approach is considerably more predictive than in the baryon
sector with nonzero helicity. This is due to the existence of an
endomorphism following from the parity transformation of the pion
couplings. We will restrict our discussion in this section to the
predictions which are common to both baryons and mesons.

We start by introducing some notation. For simplicity, we consider
$n_{f}=2$ with isospin generators $T^a$. Consider the amplitude
${\cal M}_{\alpha\beta}^a$ for pion coupling between two states
$h_\alpha(\lambda) \to h_\beta(\lambda') \pi^a$ with $h_\alpha,
h_\beta$ baryon states  with helicities $\lambda, \lambda'$,
respectively.

This amplitude is related to the matrix element of the axial
vector current between the two baryon states by PCAC as
\begin{eqnarray}\label{Mdef}
A(h_\alpha(p,\lambda) \to h_\beta(p',\lambda') \pi^a) \equiv {\cal
M}_{\alpha\beta}^a = \frac{1}{f_\pi}(p-p')_\mu \langle
h_\beta(p',\lambda') |A^{a\mu}| h_\alpha(p,\lambda)\rangle
\end{eqnarray}

The form of the pion coupling simplifies in the collinear limit,
where the baryon momenta are parallel to a common direction $\vec
p \parallel \vec p' \parallel \hat e_3$, which can be chosen for
convenience to define the $+z$ direction. In such a frame, the
pion coupling given by the expression Eq.~(\ref{Mdef}) is related
to the matrix element of a light-cone component of the axial
current, and is helicity conserving
\begin{eqnarray}\label{Xdef}
{\cal M}_{\alpha\beta}^a = \frac{1}{f_\pi}|\vec q| \langle
h_\beta(p',\lambda') |A^{a0} -A^{a3} | h_\alpha(p,\lambda)\rangle
\equiv [X^a(\lambda)]_{\beta\alpha} \delta_{\lambda\lambda'}
\end{eqnarray}
This defines the transition operator $X^a(\lambda)$ for pion
couplings to hadrons. For simplicity we will omit the $\lambda$
dependence of the operators $X^a(\lambda)$, unless explicitly
required.

The pion couplings $X^a$ and the isospin operators $T^a$ satisfy
the commutation relations
\begin{eqnarray}\label{comm1}
[T^a, T^b] = i\epsilon^{abc} T^c\,,\qquad [T^a, X^b(\lambda)] =
i\epsilon^{abc} X^c(\lambda)\,.
\end{eqnarray}
In addition to this, another commutation relation can be obtained
by considering the set of all Adler-Weisberger sum rules \cite{AW}
on baryon targets. They can be derived from an unsubtracted
dispersion relation for the $h_\alpha \pi^a \to h_\beta \pi^b$
scattering amplitude. Saturating the dispersion relation with
 one-body intermediate states, the sum rule can be expressed in operator form as
\begin{eqnarray}\label{comm2}
[X^a(\lambda), X^b(\lambda)] = i\epsilon^{abc} T^c
\end{eqnarray}
The commutation relations Eqs.~(\ref{comm1}), (\ref{comm2}) form
the Lie algebra of the SU(2)$_L \times$ SU(2)$_R$ symmetry, which
is furthermore explicitly realized on the hadronic states. This
realization may be different for each distinct helicity $\lambda$.
The resonance saturation approximation of the Adler-Weisberger sum
rule, required to derive the commutation relation
Eq.~(\ref{comm2}), can be justified, for example, by working at
leading order in the $1/N_c$ expansion.

The absence of the $\Delta I = 2$ Regge trajectories with
intersect $\alpha(0)>0$ for the pion-nucleon scattering amplitude
$h_\alpha \pi \to h_\beta \pi$ gives another commutation relation.
This connects the pion coupling operator $X^a$ with the mass
spectrum
\begin{eqnarray}\label{comm3}
[X^a(\lambda)\,, [X^b(\lambda), m^2]] \propto \delta_{ab}
\end{eqnarray}
This commutation relation implies that the baryon mass operator
has very simple transformation properties under $SU(2)_L \times
SU(2)_R$. It consists of two terms
\begin{eqnarray}\label{mass}
\hat m^2(\lambda) = \hat m_0^2(\lambda) + \hat m_4^2(\lambda)
\end{eqnarray}
with $\hat m_0^2$ transforming as a singlet $(0_L,0_R)$ under the
chiral group SU(2)$_L \times$ SU(2)$_R$, and $\hat m_4^2$
transforming as the $I=0$ component of a chiral four-vector
$(\frac12{}_{L}, \frac12{}_{R})$. In the absence of the $\hat
m_4^2$ term, the chiral symmetry would have been realized
explicitly in the mass spectrum, and baryons would be grouped into
irreducible representations of SU(2)$_L \times$ SU(2)$_R$. The
$\hat m_4^2$ term prevents such a simple picture, and implies that
in general, mass eigenstates are mixtures of different irreducible
representations of SU(2)$_L \times$ SU(2)$_R$ and do not exhibit
the pattern of degeneracies associated with a Wigner-Weyl
representation of $SU(2)_{L}\times SU(2)_{R}$. Note that $\hat
m_4^2$ is {\em not} connected with explicit chiral symmetry
breaking, which has been set to zero from the outset. Thus
Weinberg's mended symmetry scenario provides a good example of the
point emphasized in Ref.~\cite{us}, that ``restoration'' of
$SU(2)_{L}\times SU(2)_{R}$ in the spectrum requires suppression
of chirally invariant operators.

To summarize, Weinberg's dynamical arguments, based on assumed
Regge asymptotics, suffice to prove that hadrons with given
helicity $\lambda$ fall into {\em reducible} representations of
the chiral group. These representations in general may be
different for different helicities. The reducible nature of the
representations and the absence of degeneracies distinguish
Weinberg's approach from the ``chiral symmetry restoration''
picture discussed in Sec.~III. The crucial feature is the presence
of the chiral nonsinglet $\hat m_4^2$ piece in the Hamiltonian,
which connects different irreps, according to selection rules
imposed by its quantum numbers. The representation content of a
given hadron state is not fixed by these general arguments.

The mended symmetry approach, in effect, replaces the
arbitrariness of the coupling constants and masses in the
effective Lagrangian with the equally unknown representation
content of a given state. This framework is predictive only when
supplemented with a prescription for the chiral irreps content.
Once the representation content is specified, many physical
parameters (masses and couplings) are predicted in terms of just a
few input parameters. This lends such an approach considerable
predictive power. For applications of this formalism to heavy
hadron physics and exotics see
Ref.~\cite{SavB1,SavB2,SavB3,SavB4}.


We comment here on the relation of this approach to the large
$N_c$ picture of the baryons following from the work of Dashen,
Jenkins and Manohar \cite{DM,DJM}. In the large $N_c$ limit, a new
symmetry emerges in the baryon sector of QCD. This is the
contracted spin-flavor symmetry $SU(2n_f)_c$, with $n_f$ the
number of light quark flavors. The generators of the symmetry are
the spin $J^i$, isospin $I^a$, and the operator $G_0^{ia}$ defined
in terms of p-wave pion couplings to the baryons in the large
$N_c$ limit ($G^{ia} \to G_0^{ia}$)
\begin{eqnarray}
{\cal M}_{\alpha\beta}^a = \frac{1}{f_\pi} q^i N_c
[G^{ia}]_{\beta\alpha}
\end{eqnarray}
The connection to the operators introduced in Eq.~(\ref{Xdef}) is
$X^a(\lambda) \sim [G^{3a}]_{\lambda\lambda}$.

The operators $J^i, T^a, G_0^{ia}$ satisfy the commutation
relations defining the algebra of the contracted $SU(2n_f)$
symmetry
\begin{eqnarray}
&& [J^i, J^j ] = i\ \varepsilon_{ijk} J^k \,,\qquad [T^a, T^b ] = i\ \varepsilon_{abc} T^c \\
&& [ G_0^{ia}, J^j ] = i\ \varepsilon_{ijk} G_0^{ka}\,,\qquad
[G_0^{ia}, T^b] = i\ \varepsilon_{abc} G_0^{ic}\,,\qquad
[G_0^{ia}, G_0^{jb}] = 0\,.
\end{eqnarray}
Furthermore, the mass operator in the large $N_c$ limit $m_0^2$
commutes with all the generators of the contracted symmetry.

The contracted $SU(2n_f)$ symmetry relates states with different
spins and isospins, analogous to the Wigner symmetry in nuclear
physics. For example, in the symmetry limit the nucleon $N$ and
the $\Delta$ belong to the same irreducible representation of the
$SU(4)_c$, along with infinitely many other states with spin and
isospin $J=I=1/2,3/2,5/2,\cdots$. The pion couplings among these
states are fixed by the contracted symmetry, such that they are
precisely the same as the predictions of the constituent quark
model \cite{DM,DJM}.

Both the mended chiral symmetry and the large $N_c$ contracted
symmetry approach impose constraints on the form of the pion
couplings to the baryons.
In general these constraints are different, although they can
agree in special situations. The reason for the difference is the
commutation relation Eq.~(\ref{comm3}). In the large $N_c$ limit
such a commutator would vanish, since the mass operator commutes
with the pion couplings $X^a(\lambda)$. This leads to a nontrivial
chiral structure of the mass operator in the mended symmetry
approach, which does not follow from the large $N_c$ contracted
symmetry.

In certain particular cases this difference becomes immaterial,
and the predictions of the two approaches are identical. This
happens when the hadrons in a sector of given helicity $\lambda$
are assigned to one single irrep $(I_L, I_R)$ of $SU(2)_L \times
SU(2)_R$, which renders the double commutator Eq.~(\ref{comm3})
zero, in agreement with the large $N_c$ prediction. Such a
situation in illustrated in Scenario I below.

Another important difference between the two approaches lies in
their range of applicability: while the mended chiral symmetry
method is restricted to the couplings of the Goldstone bosons to
baryons, the contracted large $N_c$ symmetry is equally applicable
to the couplings of any light mesons to baryons.

We proceed to an explicit discussion of the constraints on the
mass spectrum and pion couplings from the mended chiral symmetry
approach. The  hadronic states with given helicity $\lambda$ can
be taken to transform as linear combinations of irreducible
representations $(j_L, j_R)$ of SU(2)$_L \times$ SU(2)$_R$
\begin{eqnarray}\label{Mlambda}
|h(\lambda)\> \sim \sum_{j_L, j_R} c_{j_L j_R}(\lambda) |j_{L},
j_{R}\>
\end{eqnarray}
where we have suppressed all labels on the states except for their
$SU(2)_{R}\times SU(2)_{L}$ representation content. In general,
the mass operator $\hat m_4^2$ will mix different representations
(barring accidental degeneracies), such that the states
$|h(\lambda)\rangle $ will be nondegenerate, unless the sum over
representations contains only one term.

Parity invariance relates the representation contents of hadronic
states with opposite helicity.\footnote{Of course Lorentz
invariance constrains the representation content so that states of
the same hadron with different $|\lambda|$ are degenerate.} Acting
with the parity operator $\hat P$ on a hadron state $h_\alpha(\vec
p, \lambda)$ gives a state of opposite helicity and momentum $\hat
P | h_\alpha(\vec p, \lambda)\rangle = \pi_h | h_\alpha(-\vec p,
-\lambda)\rangle$, with $\pi_h$ the intrinsic parity of the state
$h$. It is more convenient to consider the \emph{normality}
operator $\Pi = \hat P R_\pi^y$, the product of parity and a
$180^\circ$ rotation, which restores the momenta to their original
direction. The action of this operator on the hadron states is
$\Pi | h_\alpha(\vec p, \lambda)\rangle = \pi_h (-)^{j_h} |
h_\alpha(\vec p, -\lambda)\rangle$.

The generators of the symmetry transform under $\Pi$ as
\begin{eqnarray}\label{XTparity}
\Pi X^a(\lambda) \Pi = - X^a(-\lambda)\,, \qquad \Pi T^a \Pi = T^a
\end{eqnarray}
The mass operator commutes with the normality operator. Using this
relation, it is easy to see that the states $|h(-\lambda)\>$
furnish also a reducible representation of the chiral algebra,
containing the representations $(j_R, j_L)$ obtained from those in
Eq.~(\ref{Mlambda}) by exchanging $j_L \leftrightarrow j_R$
\begin{eqnarray}\label{parity}
|h(-\lambda)\> \sim \sum_{j_L, j_R} c_{j_R j_L}(\lambda) |j_L,
j_R\>
\end{eqnarray}

To illustrate the application of this approach, we consider two
particular cases. The first scenario describes a $N,\Delta$ pair
with the same mass, which reproduces the $NN\pi, N\Delta\pi,
\Delta\Delta\pi$ couplings of the quark model. The second scenario
describes a pair of spin 1/2 nucleons of opposite parity.

\paragraph {Scenario I}

 A model for a nonstrange nucleon and delta baryons can be constructed
by placing the hadrons into minimal chiral multiplets. This model
was first proposed in Ref.~\cite{Weinberg:1969hw} and discarded
because it has the unphysical feature that the nucleon and delta
are degenerate. It was proposed again in Ref.~\cite{Wtalk} as a
possible justification for the quark model with SU(4) spin-flavor
symmetry. This scenario comes closest to the large $N_c$ picture
of the ground state baryons, where the nucleon-$\Delta$ mass
splitting is of order $1/N_c$.

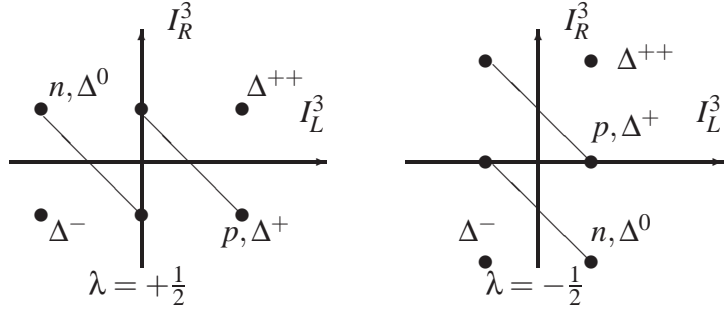
\begin{figure}\label{figweight}
\begin{picture}(300,100)(0,0)
\linethickness{1pt} \put(0,50){\vector(1,0){120}}
\put(50,10){\vector(0,1){90}} \put(12,30){\circle*{5}}
\put(50,30){\circle*{5}} \put(88,30){\circle*{5}}
\put(12,70){\circle*{5}} \put(50,70){\circle*{5}}
\put(88,70){\circle*{5}} \put(50,30){\line(-1,1){37}}
\put(88,30){\line(-1,1){37}} \put(110,65){$I_L^3$}
\put(60,100){$I_R^3$} \put(30,0){$\lambda = +\frac12$}
\put(90,75){$\Delta^{++}$} \put(15,75){$n,\Delta^{0}$}
\put(15,20){$\Delta^{-}$} \put(80,20){$p,\Delta^{+}$}
\put(150,50){\vector(1,0){120}} \put(200,10){\vector(0,1){90}}
\put(180,12){\circle*{5}} \put(180,50){\circle*{5}}
\put(180,88){\circle*{5}} \put(220,12){\circle*{5}}
\put(220,50){\circle*{5}} \put(220,88){\circle*{5}}
\put(220,50){\line(-1,1){37}} \put(220,12){\line(-1,1){37}}
\put(260,65){$I_L^3$} \put(210,100){$I_R^3$}
\put(180,0){$\lambda=-\frac12$} \put(230,85){$\Delta^{++}$}
\put(220,60){$p,\Delta^{+}$} \put(220,20){$n,\Delta^{0}$}
\put(170,20){$\Delta^{-}$}
\end{picture}
\caption{\small  Weight diagrams for the chiral representations
$(1,\frac12)$ and $(\frac12,1)$ for states with helicity $\lambda
= +\frac12$ (left) and $\lambda = - \frac12$ (right),
respectively. The states on the same diagonal have the same third
component of the isospin $I^3=\frac12(I_L^3+I_R^3)$, and
correspond to the nucleons and $\Delta$ states shown. }
\end{figure}

We place each helicity sector into irreducible representations of
SU(2)$_L \times$ SU(2)$_R$
\begin{eqnarray}
\lambda = +\frac12 &:& (1_L, \frac12{\,}_{R}) \ \supset\  N, \Delta \\
\lambda = + \frac32 &:& (\frac32{\,}_{L}, 0_R) \ \supset\  \Delta
\end{eqnarray}
The transformation properties of the negative helicity states are
obtained from these via the parity relation Eq.~(\ref{parity}) and
are
\begin{eqnarray}
\lambda = -\frac12 &:& (\frac12{\,}_{L}, 1_R) \ \supset\  N, \Delta \\
\lambda = - \frac32 &:& (0_L, \frac32{\,}_{R}) \ \supset\  \Delta
\end{eqnarray}
We show in Figure \ref{figweight} the weight diagrams of the
chiral representations corresponding to $\lambda = \pm \frac12$,
and the associated hadron states.

The mass matrix (see eq.~(\ref{mass})) for the helicity
$\lambda=+1/2$ is proportional to the unit matrix in the basis
$(N_{+1/2}, \Delta_{+1/2})$, since the $\hat m_4^2$ term does not
have diagonal matrix elements on a irreducible  representation
$(I_L,I_R)$. Thus the nucleon and delta are degenerate in this
model.

Collinear pion emission connects only states with the same
helicity, and the corresponding axial matrix elements can be
obtained by expressing the axial current in terms of the
generators of the chiral group $X^a = \frac12(I_R^a - I_L^a)$.
Defining the reduced matrix elements for p-wave pion emission as
\begin{eqnarray}
\langle J'I'; \lambda I'_3 | X^a |JI; \lambda I_3 \rangle =
\frac{1}{2J'+1} X(J',J) \langle J'\lambda|J1;\lambda 0\rangle
\langle I'I'_3|I1; I_3 a\rangle
\end{eqnarray}
one finds $X(1/2,1/2) = -5/\sqrt2,\ X(1/2,3/2) = -4,\
X(3/2,3/2)=-5\sqrt2$, in agreement with the predictions of the
quark model with SU(4) spin-flavor  Wigner  symmetry. In
particular, the result for $X(1/2,1/2)$ is equivalent to $g_A/g_V
= 5/3$.
 The agreement with the experimental result $g_{A}/g_{V}=1.2695 \pm 0.0029$ can be improved
by refining the model and adding a second chiral representation
$(\frac12 {\,}_{L},0_R)$ in the $\lambda=+\frac12$ sector. The
mixing angle of the two representations is chosen appropriately
such as to reproduce the data on $g_A/g_V$ \cite{Weinberg:1969hw}.
We do not consider this further, and refer the reader to
Weinberg's paper for details \cite{Weinberg:1969hw}.

\paragraph{Scenario II}

Consider a different model, containing the two chiral
representations $(0,\frac12)$ and $(\frac12, 0)$. In the helicity
$+1/2$ sector, there are two spin-1/2 nucleons $N_A, N_B$ which
are linear combinations of the two chiral representations, with a
mixing angle $\theta$
\begin{eqnarray}
\lambda = +\frac12 &:& N_A = \cos\theta (0_L, \frac12{\,}_{R}) + \sin\theta (\frac12{\,}_{L}, 0_R)  \\
& & N_B = -\sin\theta (0_L, \frac12{\,}_{R}) + \cos\theta
(\frac12{\,}_{L}, 0_R)
\end{eqnarray}
We take these states to be mass eigenstates, with eigenvalues
$m_A, m_B$.  They are related to the chiral singlet and chiral
quartet terms as follows. Consider the squared mass matrix in the
basis of the irreducible representations
 $(\frac12, 0)$ and $(0,\frac12)$
\begin{eqnarray}
m^{2}_{\lambda = +1/2} = \left(
\begin{array}{cc}
\mu^{2}_L & \alpha \\
\alpha & \mu^{2}_R \\
\end{array}
\right)
\end{eqnarray}
where $\mu^{2}_{L,R}$ are the diagonal matrix elements of the
chiral singlet $\hat m^{2}_0$, and $\alpha$ is the off-diagonal
matrix element of the chiral quartet term in the mass-squared
operator of eq.~(\ref{mass}). Their relation to the hadronic
parameters is
\begin{eqnarray}
 \mu^{2}_L = m^{2}_A \sin^2 \theta + m^{2}_B \cos^2\theta \,,\,\,
\mu^{2}_R = m^{2}_A \cos^2 \theta + m^{2}_B \sin^2\theta\,,\,\,
\alpha = \frac12 (m^{2}_A - m^{2}_B) \sin 2\theta
\end{eqnarray}

The helicity $-1/2$ sector can be constructed from these states by
applying the normality transformation, as explained above.
Choosing the two nucleons to have opposite parities $\pi_{N_A} =
+1, \pi_{N_B} = -1$, the representation content of the helicity
$-\frac12$ states is
\begin{eqnarray}
\lambda = -\frac12 &:& \Pi N_A =
\sin\theta (0_L, \frac12{\,}_{R}) + \cos\theta (\frac12{\,}_{L}, 0_R)  \\
& & \Pi N_B = \cos\theta (0_L, \frac12{\,}_{R}) - \sin\theta
(\frac12{\,}_{L}, 0_R)
\end{eqnarray}

Next consider the couplings of these states to pions. The pion
coupling operator $X^a = \frac12 (I_R^a - I_L^a)$ is related to
the generators of the symmetry, and thus couples only states in
the same chiral representations. The simplest way to derive the
pion couplings is to consider the transitions $p_{i\uparrow } \to
n_{j\uparrow} \pi^+$, which are mediated by the operator $X^- =
X^1 - iX^2$. For example, the diagonal pion transition $ N_A\to
N_A \pi$ is computed as
\begin{eqnarray}
&& A(p_{A\uparrow } \to n_{A\uparrow} \pi^+) \sim \\
&&  \quad \Big( \cos\theta \langle (1,0)_L,
(\frac12,-\frac12){\,}_{R} | +
\sin\theta \langle (\frac12,-\frac12){\,}_{L}, (1,0){\,}_R  | \Big) \nonumber \\
&& \quad \times \,\, X^- \Big( \cos\theta |(1,0){\,}_L,
(\frac12,+\frac12){\,}_{R}\rangle +
\sin\theta |(\frac12,+\frac12){\,}_{L}, (1,0){\,}_R \rangle \Big) \nonumber \\
&& \hspace{-1cm} = \cos^2\theta \langle (\frac12,-\frac12){\,}_{R}
| I_R^- | (\frac12,+\frac12){\,}_{R} \rangle - \sin^2\theta
\langle (\frac12,-\frac12){\,}_L | I_L^- |
(\frac12,+\frac12){\,}_L \rangle = \cos 2\theta \nonumber
\end{eqnarray}
where in the last line we show only the chiral isospin (and its
third component) changed by the $I_{L,R}^-$ operator. According to
the effective Lagrangian of eq.~(\ref{Lnl}), this amplitude is
proportional to the coupling $c_2 + c_4$. In a similar way one can
obtain all the other couplings in this effective Lagrangian, with
the results
\begin{eqnarray}
&& c_2 + c_4 \sim \cos 2\theta \\
&& c_2 - c_4 \sim - \cos 2\theta \\
&& 1 - c_3 \sim - \sin 2\theta
\end{eqnarray}

This illustrates how a special chiral representation content in
the framework of the mended chiral symmetry can lead to  specific
relations among pion couplings in the effective Lagrangian. In
particular, choosing $\theta = 3\pi/4$ reproduces the predictions
of the chiral doubling model with linearly transforming baryons
($c_{2-4} = 0$), but without mass degeneracy. The mass parameters
of this model are $\mu^{2}_L = \mu^{2}_R = (m^{2}_A + m^{2}_B)/2$
and $\alpha = -(m^{2}_A-m^{2}_B)/2$, with nondegenerate baryons
$m^{2}_A - m^{2}_B \neq 0$.

Of course, any such explanation leaves open the dynamical origin
of this (or any other) particular value of the mixing angle, as
well as the representation content. More complicated
representation contents are also possible, e.g. including also
$(I_L, I_R) = (1,1/2), (1/2,1)$ as in Scenario 1. Such questions
could be addressed for example using lattice QCD methods, as
discussed recently in  Ref.~\cite{gA}.


\section{ Conclusion  and Discussion}

Although parity doubling was observed in the baryon spectrum more
than 35 years ago, a complete understanding of its dynamical
origin is still lacking. In this paper we both evaluate the
empirical evidence for parity doubling and consider some possible
theoretical explanations.  We propose a new mechanism, based on
the dynamical suppression of the $U(1)_A$ violation, and point out
a prediction that distinguishes it from $SU(2)_{L}\times
SU(2)_{R}$ restoration.

Most of the evidence previously offered in favor of parity
doubling has been obtained by visually grouping states into
multiplets. Given the number of broad, closely spaced states in
each channel, this procedure is both ambiguous and imprecise. It
does not take into account the resonance widths, nor the fact that
different resonances are established with varying degrees of
certainty. We propose in this paper a quantitative measure of
parity doubling, which is free of these problems. First, we
introduce an overlap integral $\Omega_{IJS} (\{ C\})$  of the
spectral densities in channels with the same spin $J$, isospin $I$
and strangeness $S$, but opposite parity. This is computed on an
ensemble of control samples (each of them denoted by $\{ C \}$),
obtained from the real world by arbitrarily permuting the parities
of the hadronic states. Finally, the value of the overlap integral
$\Omega_{IJS}$ in the real world is compared against the
statistical distribution on the ensemble, which assigns a
probability measure ($p$-value) to the statement of parity
doubling.

The results of our study gives strong evidence for parity doubling
in the nonstrange baryon sector (both $N$ and $\Delta$), with
spins from $J=1/2$ to $7/2$. For larger spins the number of states
decreases, and we do not expect the control samples to be
representative statistically. Still, there is good evidence for
parity doubling among high spin, non-strange baryons as can be
seen from  Table~\ref{table1}. On the other hand, the correlations
between positive and negative parity \emph{strange} baryons are
not significant. However, some of the states with strangeness
expected in the quark model have not yet been seen, which could
partly be responsible for the weaker correlations observed in
these sectors. A new experimental attack on baryon spectroscopy
could shed significant light on the systematics and origins of
parity doubling.

In principle our statistical test could be used to search for more
complex correlations, such as those expected in models of ``chiral
symmetry restoration''. In such schemes, additional degeneracies
involving hadrons of different  isospins (but the same spin)  are
expected on top of parity doubling. The pattern of the mass
spectrum in this case depends on the representation,
$(j_{L},j_{R})$, of $SU(2)_L \times SU(2)_R$ assumed to be
realized. For example, in the $(1/2, 1) \oplus (1, 1/2)$ scheme of
Ref.~\cite{quartet} one expects the baryons to be grouped into
``quartets'' of the form $(N^+_J, N^-_J, \Delta^+_J, \Delta^-_J)$
with the same spin $J$. We have not used our statistical method to
look for more complicated correlations.  This would require higher
dimensional integrals analogous to eq.~(\ref{statistic}).  In any
case no general correlations of this form are observed by eye.
However, it is possible that more complicated chiral
representations may not appear uniformly in the spectrum, but be
localized in some mass region. This would render our method
inapplicable, due to the sparseness of the corresponding control
sample.

We have examined a number of possible explanations of parity
doubling.  First we have tried to clarify the relationship between
the usual $SU(2)_{L}\times SU(2)_{R}$ chiral symmetry of QCD and
parity doubling.  The terminology used in the literature
discussion has been confusing.  In an important sense (for
massless $u$ and $d$ quarks) $SU(2)_{L}\times SU(2)_{R}$ is not
broken at all, but instead represented in the Nambu-Goldstone form
by massless pions and isospin multiplets of hadrons that transform
non-linearly under chiral transformations.  There is nothing to
``restore''. Instead, we frame the question as follows:  ``What is
needed in order that baryons fall into Wigner-Weyl representations
of $SU(2)\times SU(2)$with the usual degeneracies and relations
among couplings''? The answer, presented in Ref.~\cite{us} and
reviewed here, is that a set of $SU(2)_{L}\times SU(2)_{R}$
\emph{invariant} operator matrix elements must be suppressed,
basically amounting to decoupling the pion from the parity doublet
states.  If this decoupling were complete, it would, in effect,
leave \emph{two independent} $SU(2)_{L}\times SU(2)_{R}$
symmetries:  one under which the pion and non-parity-doubled
states transform \emph{non-linearly}, and the other under which
the parity doubled states transform \emph{linearly}.  A toy model
of this type has recently been proposed by Cohen and
Glozman~\cite{cgnew}.  The question is what dynamics are
responsible for suppressing these operator matrix elements and
giving rise to this additional symmetry? If the dynamics behind
effective chiral symmetry restoration applies generically to
hadrons high in the spectrum, as advocated, for example, in
Ref.~\cite{Glozman:2004gk}, then parity doubling should be
universal in that domain.  It would be puzzling if, for example,
some ({\it e.g.\/}) highly excited mesons are parity doubled, but
some are not.  A search for such exceptions would be a
particularly effective way to test this explanation of parity
doubling.

We emphasize that parity doubling alone is not conclusive evidence
for effective chiral symmetry restoration.  As discussed in
Sec.~3A, chiral symmetry realized linearly on a set of baryons
requires that pions decouple from the parity doubled states.  In
addition, one would have to check the validity of certain
relations for axial couplings between the members of the putative
chiral multiplet implied by the symmetry algebra implemented in
the Wigner-Weyl mode.

 Next we considered the possibility that parity doubling
is a consequence of dynamical suppression of $U(1)_A$ symmetry
breaking in a subsector ${\mathfrak h}$ of the Hilbert space of
QCD. Note that this is not equivalent to $U(1)_A$ restoration, as
realized for example in large $N_c$ QCD. For example, the effects
of the anomaly are still present, and give the $\eta,\eta'$ the
masses observed in nature. Rather, we require only that the matrix
elements of the divergence of the axial singlet current are very
small on a subset of baryon states $\langle B_1 |\p_\mu j_5^\mu |
B_2\rangle \approx 0$, with $B_1, B_2 \in {\mathfrak h}$. This
condition is satisfied if the matrix elements of the gluon anomaly
term taken between non-strange baryons of opposite parity are very
small. The hypothesis that the matrix elements of $\tilde FF$ are
suppressed among excited baryons may be amenable to study using
lattice QCD methods.

Although it predicts parity doublets, the mechanism of dynamical
suppression of $U(1)_A$ symmetry breaking does not make
predictions for the pion couplings to the doublet states (except
that the pions must decouple from parity doubled states at zero
momentum, as required by the Goldberger-Treiman relation). Such
relations would be obtained, for example, if we required the
doublet to be left invariant under $U(1)_A$ and parity
transformations.

In addition to suppression of $U(1)_{A}$ symmetry violation, we
review two other possible origins for parity doubling. First we
examine arguments recently put forward that parity doubling in the
spectrum can be related to chiral symmetry restoration at short
distances, a phenomenon well grounded in perturbative QCD.  The
only input from short distance chiral symmetry is the
superconvergence relation, eq.~(\ref{super}).  The rest of the
argument requires input from phenomenological models ({\it e.g.\/}
linear excitation spectra) and large $N_{c}$ implemented in an
unusual way --- examining the large $n$ (excitation) limit, but
taking $N_{c}\to\infty$ first.  ${\cal O}(1/N_{c})$ corrections
seem capable of invalidating the conclusions.  In any event this
exercise provides an example of the extent of additional dynamics
required to obtain effective chiral symmetry restoration. Finally
we look at the possibility that parity doubling may be a
consequence of an intrinsic deformed state that spontaneously
violates parity. A mechanism familiar from nuclear and molecular
physics, this leads one to expect parity doubling at high spin for
both strange and non-strange baryons, but gives little insight
into why spin 1/2 and 3/2 nucleons and $\Delta$s should show
statistically significant evidence for the symmetry.

An alternative way of obtaining relations among masses and axial
couplings which adds dynamical input in a minimal way is offered
by the mended chiral symmetry approach of Weinberg
\cite{Weinberg:1969hw,Weinberg:1990xn}. In this approach the set
of all pion couplings among baryon states $X^a$ (in sectors of
given helicity), together with the isospin operators $T^{a}$, are
shown to satisfy the $SU(2)_{L}\times SU(2)_{R}$ algebra. However,
input from Regge theory shows that the mass operator is not a
singlet under this symmetry, but contains in addition also a
chiral 4-vector term.

This implies that the hadrons transform in {\em reducible}
representations of the $SU(2)_{{L}}\times SU(2)_{R}$ symmetry.
Models of hadrons can be constructed by postulating a
representation content, and the corresponding mixing angles. Even
minimal models containing one or two such representations can
reproduce the properties of the lowest lying baryons remarkably
well \cite{Weinberg:1969hw,Weinberg:1990xn}.

As examples of this approach we presented two such models. First,
to introduce the ideas we describe a model containing a
$(N_{1/2},\Delta_{3/2})$ degenerate pair of the same parity, which
reproduces the quark model predictions for pion couplings. This
model was previously considered in Refs.~\cite{Wtalk,SavB2}. The
second  model is new and contains two nucleons of opposite parity
and spins $J=1/2$. Their masses and couplings are determined by
just three mass parameters and one mixing angle among chiral
representations.  We  show that, for a particular value of the
mixing angle, the pion couplings to the parity doublet take
precisely the same values as those expected if chiral symmetry
were realized linearly on the two states,  but without mass
degeneracy in the spectrum. This illustrates a natural realization
of such relations which does not require chiral symmetry
restoration.

\section{Acknowledgments}

 We thank W.~Bardeen, S.~Beane, D.~Bugg, T.~Cohen, M.~Golterman, L.~Glozman, A.~Manohar,  M.~Shifman
and A.~Vainshtein for conversations and correspondence relating to
this work. This work is supported in part by funds provided by the
U.S.~Department of Energy (D.O.E.) under cooperative research
agreement DE-FC02-94ER40818.

\bibliographystyle{aipprocl}   

\begin{thebibliography}{99}
\bibitem{duck} To our knowledge, the possibility that parity
doubling in QCD is a consequence of the suppression of $U(1)_{A}$
violating matrix elements between baryons was first suggested in
\cite{duck2}.
\end{thebibliography}


\end{document}